\documentclass[12pt]{iopart}

\usepackage{iopams}
\usepackage{graphicx}
\usepackage{subfig}
\usepackage{color}
\usepackage{ulem}
\usepackage{color}
\usepackage{ulem}
\usepackage{hyperref}

\begin{document}
	
	\title{Species dependence of the impurity injection induced poloidal flow and magnetic island rotation in a tokamak}
	
	\author{Shiyong Zeng}
	\address{Department of Engineering and Applied Physics, University of Science and Technology of China, Hefei, Anhui 230026, China}
	
	\author{Ping Zhu}
	\address{International Joint Research Laboratory of Magnetic Confinement Fusion and Plasma Physics, State Key Laboratory of Advanced Electromagnetic Engineering and Technology, School of Electrical and Electronic Engineering, Huazhong University of Science and Technology, Wuhan, Hubei 430074, China}
	\address{Department of Engineering Physics, University of Wisconsin-Madison, Madison, Wisconsin 53706, USA}
	\ead{zhup@hust.edu.cn}
	
	\author{Haijun Ren}
	\address{Department of Engineering and Applied Physics, University of Science and Technology of China, Hefei, Anhui 230026, China}
	\ead{hjren@ustc.edu.cn}

	\title[Impurity injection induced poloidal flow and island rotation in a tokamak]{}
	
	\newpage
	\begin{abstract}
		\label{Sec:abstract}
		Recent experiments have demonstrated the species dependence of the impurity poloidal drift direction along with the magnetic island rotation in the poloidal plane. Our resistive MHD simulations have reproduced such a dependence of the impurity poloidal flow, which is found mainly determined by a local plasmoid formation due to the impurity injection. The synchronized magnetic island rotation is dominantly driven by the electromagnetic torque produced by the impurity radiation primarily through the modification to the axisymmetric components of current density.

	\end{abstract}

	\section{Introduction}
	\label{Sec: Introduction}	
	The impurity control for steady-state tokamak operation and its injection for disruption mitigation have been subjects of great interests and investigations over the past decades.
	Recent J-TEXT massive gas injection (MGI) experiments demonstrate that high-Z impurity (Argon) density injected from the bottom of the device only prefer to drift to the magnetic high field side (HFS) along the poloidal direction in the poloidal plane, whereas the low-Z impurity (Helium) density tends to drift towards the magnetic low field side (LFS) firstly during the pre-thermal quench (pre-TQ) phase and then reversely drifts to the HFS in the subsequent thermal quench (TQ) phase for the same initial conditions \cite{Li_2020}. Such a difference in the poloidal drift direction between low-Z and high-Z impurity species would disappear in the case of lower valve voltage, i.e. with weaker impurity injection and penetration. Similar poloidal reversal is also found in the magnetic island rotation along with the impurity flow in the case of low-Z impurity injection \cite{Li_2020}. The underlying mechanism for such an impurity species dependence of the poloidal rotation direction during the impurity injection process observed in experiments had not been well understood until recently.

	The impurity induced tearing mode (TM) is widely observed in experiments, and most studies focus on the island growth affected by the impurity radiation \cite{Rebut1985,Rutherford1985,White1977,Xu_2017,Gates2012,Zeng_2023_tm}. On the one hand, the impurity radiation modifies the pressure profile locally to change the Pfirsch-Schl\"{u}ter current or the bootstrap current; on the other hand, it affects the current profile directly through the temperature dependent plasma resistivity. Both influence the tearing instability growth.
	Meanwhile, how the magnetic island rotation may be directly affected by the impurity radiation is less clear \cite{XuMing2022}.	
	Previous NIMROD simulations show the helical expansion of Ne impurity away from the injection location only towards the HFS, and a simple magnetic nozzle model is used to explain the parallel spreading of the asymmetric impurity flow, which enables the growth of impurity flow spread along the field lines towards the region with stronger increasing magnetic field \cite{Izzo_2015}. However, this model cannot explain the species dependence of the poloidal impurity flow. Nor can the $E\times B$ or $\nabla B$ drift account for the poloidal flow of different impurity species injected from the same location observed on J-TEXT, since these drifts are all towards the same poloidal direction regardless of impurity ion species.
	

	On the other hand, the pellet fueling is similar to the impurity injection process despite the former using the same particle species as the plasma.
	ASDEX-Upgrade tokamak has demonstrated a high-efficiency pellet fueling from the HFS in contrast to the LFS, which is correlated to a local enhanced-$\beta$ plasmoid formation \cite{Lang1997PRL,Muller1999PRL}. An enhanced core density assimilation is found as well when using mixed H$_2$ + Ne pellet injection for the disruption mitigation due to the suppressed plasmoid formation by the increased Ne radiation \cite{Matsuyama2022PRL}.
	A pellet injection model is developed to describe the cross-field drift of ionized pellet ablation matter in tokamak plasma, which includes the effect of pressure variation, curvature and magnetic shear \cite{Parks2005PRL}. We find this model capable of explaining our simulation results well, suggesting some common physics underlying the gas and pellet injection processes.

	In this work, we use 2D simulations to demonstrate the interaction between the impurity and plasma, and how the formation of a local enhanced-$\beta$ plasmoid due to the low-Z impurity (He) injection leads to the different direction of density poloidal flow, which are consistent with the experimental observations. Then, we use 3D simulation results to show how the impurity injection induced poloidal flow and magnetic island are primarily driven by the electromagnetic force to rotate synchronously in the poloidal plane.

	The rest of this paper is arranged as follows: Section \ref{Sec: model} introduce the simulation model and setup. Section \ref{Sec: 2D cases} shows the 2D simulation results on the interaction between the impurity and the plasma in absence of the tearing mode. Section \ref{Sec: 3D cases} reports the 3D simulation results showing that the impurity injection drives poloidal flow and the magnetic island rotation in the poloidal plane. Discussion and summary are made in Section \ref{Sec:summary}.



	
	\section{NIMROD/KPRAD model and simulation setup}
	\label{Sec: model}
	Our simulations in this work are based on the single-fluid resistive MHD model implemented in the NIMROD code \cite{Sovinec2004}, and a simplified module for impurity radiation adapted from the KPRAD code \cite{KPRAD}. The equations for the impurity-MHD model are as follows
	
	\begin{eqnarray}
		\rho \frac{d\vec{V}}{dt} = - \nabla p + \vec{J} \times \vec{B} + \nabla \cdot (\rho \nu \nabla \vec{V})
		\label{eq:momentum}
		\\
		\frac{d n_i}{dt} + n_i \nabla \cdot \vec{V} = \nabla \cdot (D \nabla n_i) + S_{ion/3-body}
		\label{eq:contiune2}
		\\
		\frac{d n_{Z,Z=0-18}}{dt} + n_Z \nabla \cdot \vec{V} = \nabla \cdot (D \nabla n_Z) + S_{ion/rec}
		\label{eq:contiune3}
		\\
		n_e \frac{d T_e}{dt} = (\gamma - 1)[n_e T_e \nabla \cdot \vec{V} + \nabla \cdot \vec{q_e} - Q_{loss}]
		\label{eq:temperature}
		\\
		\vec{q}_e = -n_e[\kappa_{\parallel} \hat{b} \hat{b} + \kappa_{\perp} (\mathcal{I} - \hat{b} \hat{b})] \cdot \nabla T_e
		\label{eq:heat_flux}
		\\
		\vec{E} + \vec{V} \times \vec{B} = \eta \vec{j}
		\label{eq:ohm}
	\end{eqnarray}
	Here, $n_i$, $n_e$, and $n_Z$ are the main ion, electron, and impurity ion number density respectively, and $Z$ is the charge number, $\rho$, $\vec{V}$, $\vec{J}$, and $p$ the plasma mass density, velocity, current density, and pressure respectively. $T_e$ and $\vec{q}_e$ are the electron temperature and heat flux respectively. $D$, $\nu$, $\eta$, and $\kappa_{\parallel} (\kappa_{\perp})$ are the plasma diffusivity, kinematic viscosity, resistivity, and parallel (perpendicular) thermal conductivity respectively, $\gamma$ the adiabatic index, $S_{ion/rec}$ the density source from ionization and recombination, $S_{ion/3-body}$ the contribution from the 3-body recombination, $Q_{loss}$ the energy power loss, $\vec{E} (\vec{B})$ the electric (magnetic) field, $\hat{b}=\vec{B}/B$, and $\mathcal{I}$ the unit dyadic tensor.

\begin{figure}[ht]
		\begin{center}
			\includegraphics[width=0.48\textwidth,height=0.3\textheight]{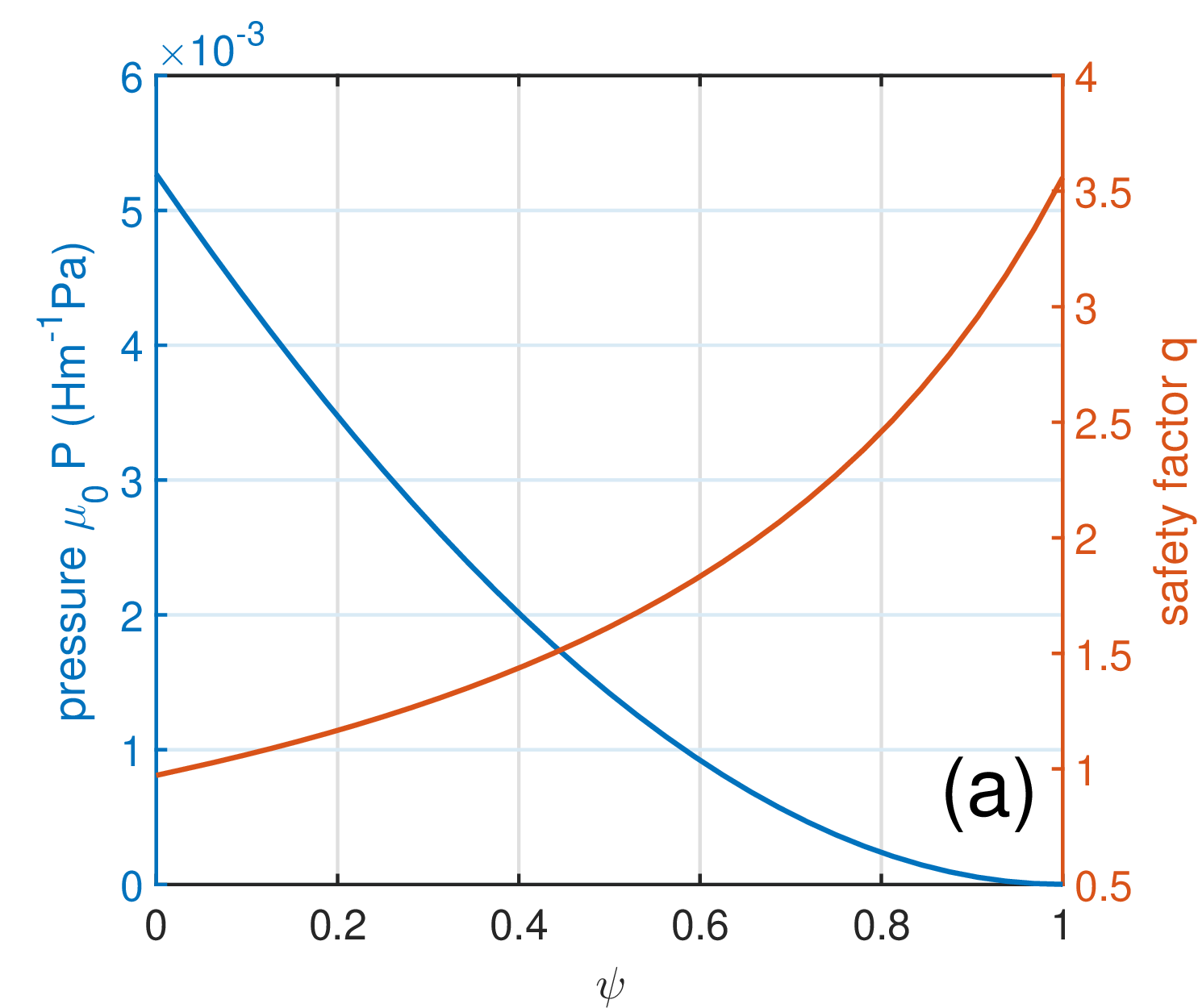}
			\includegraphics[width=0.48\textwidth,height=0.3\textheight]{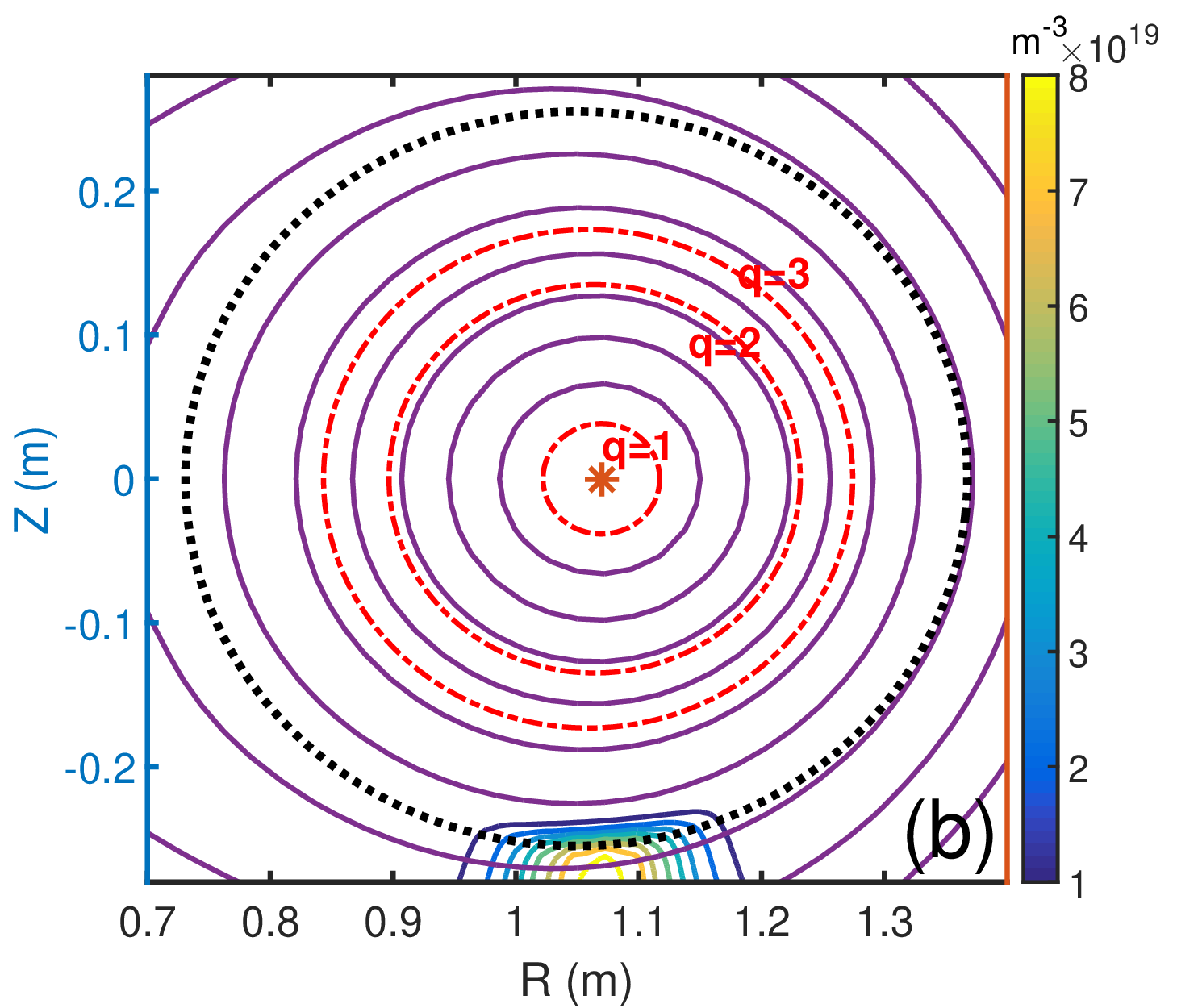}
		\end{center}
		\caption{(a) Initial equilibrium pressure $p$ and safety factor $q$ as functions of the normalized flux function $\psi$. (b) Contours of the initial equilibrium magnetic flux $\psi$ (purple solid lines) and impurity density injection (flushed color, color bar in unit $m^{-3}$) in the poloidal plane, where the simulation domain boundary is denoted as the black dashed-line, and equilibrium $q=1,2,3$ rational surfaces are denoted as the red dashed-lines.}
		\label{fig:equilibrium}
	\end{figure}

	An initially static tokamak equilibrium with a circular shaped boundary is considered in this work for simplicity, in order to exclude the potential effects from poloidal shaping and plasma rotation.
	The major radius $R_0=1.05m$ and the minor radius $a=0.25m$, the plasma central toroidal magnetic field $B_{t0}=1.75T$ and the total plasma current $I_p=150kA$, the core electron density $n_{e0}=1.875 \times 10^{19} m^{-3}$ and electron temperature $T_{e0}=700eV$, the central safety factor $q_0=0.955$ and edge safety factor $q_a=3.797$, the equilibrium profiles are shown in Fig. \ref{fig:equilibrium}(a) and are adopted from typical J-TEXT experimental parameters \cite{Wang_2022}.
	The core Lundquist number $S = 3.847\times10^8 $ according to the core electron temperature $T_{e0}$, and the magnetic Prandtl number $P_{rm}= 6.616\times 10^3$ \cite{ItohJPSJ1993}. A temperature dependent anisotropic thermal conductivity model is adopted with the perpendicular component $\kappa_{\perp}=(700/T_e)^{1/2}(1/B^2) m^2/s$ and the parallel component $\kappa_{\parallel}=10^6(T_e/700)^{5/2} m^2/s$ \cite{Braginskii1965}. A constant diffusivity $D=2 m^2/s$ is prescribed for all particle species, which is close to the experimental measurement value \cite{Delgado-Aparicio_2011}.
	The Spizter resistivity model $\eta \sim T_e^{-3/2}$ is adopted and the core resistivity is $\eta_0= 5.129\times 10^{-9} \Omega \cdot m$. A constant kinematic viscosity $\nu= 27 m^2/s$ is used and found numerically stabilizing and convergent.
	The impurity injection from the bottom edge of the plasma is modeled as an initially static density deposition localized in the poloidal plane and its peak value is approximate $4$ times higher than the core electron density (Fig. \ref{fig:equilibrium}b),
	\begin{equation}
		S_{imp}=n_{imp} \left [ 100 \tanh{ \left ( \frac{r}{r_v}-1 \right ) } +1 \right ]  \exp{ \left [ -\left ( \frac{\theta - \theta_0}{15} \right ) ^2  -\left ( \frac{\phi - \phi_0}{15} \right ) ^2 \right ] }
		\label{imp}
	\end{equation}
	which is the only perturbation introduced to the initial equilibrium. Here $n_{imp}$ is the injected impurity density, $r_v$ the radius of plasma boundary, and $\theta_0$ ($\phi_0$) the poloidal (toroidal) angle of the impurity gas injection location.

	\section{2D simulations of impurity-plasma interaction}
	\label{Sec: 2D cases}
	
\begin{figure}[ht]
		\begin{center}
			\includegraphics[width=0.85\textwidth,height=0.35\textheight]{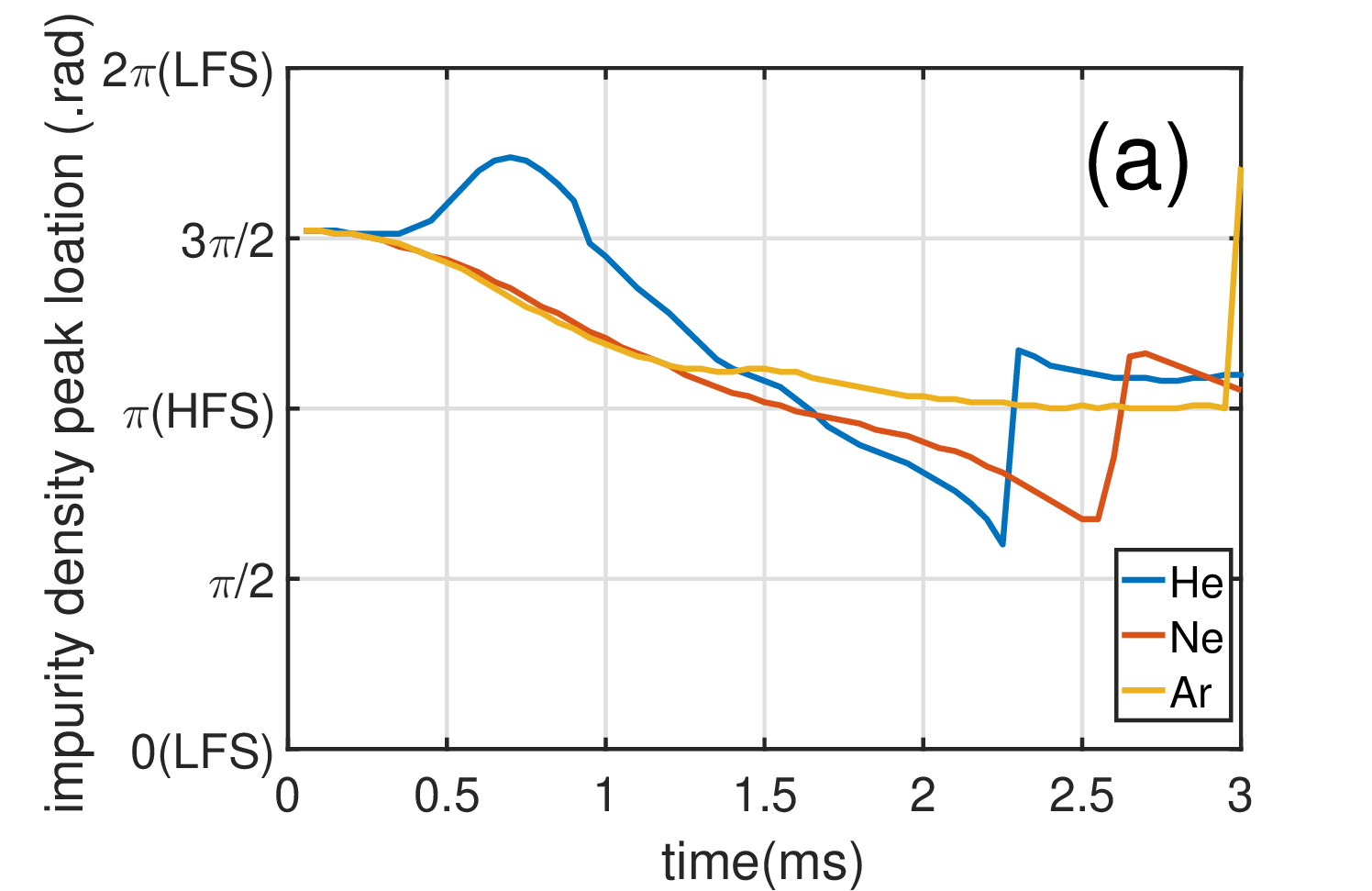}
			\includegraphics[width=0.85\textwidth,height=0.35\textheight]{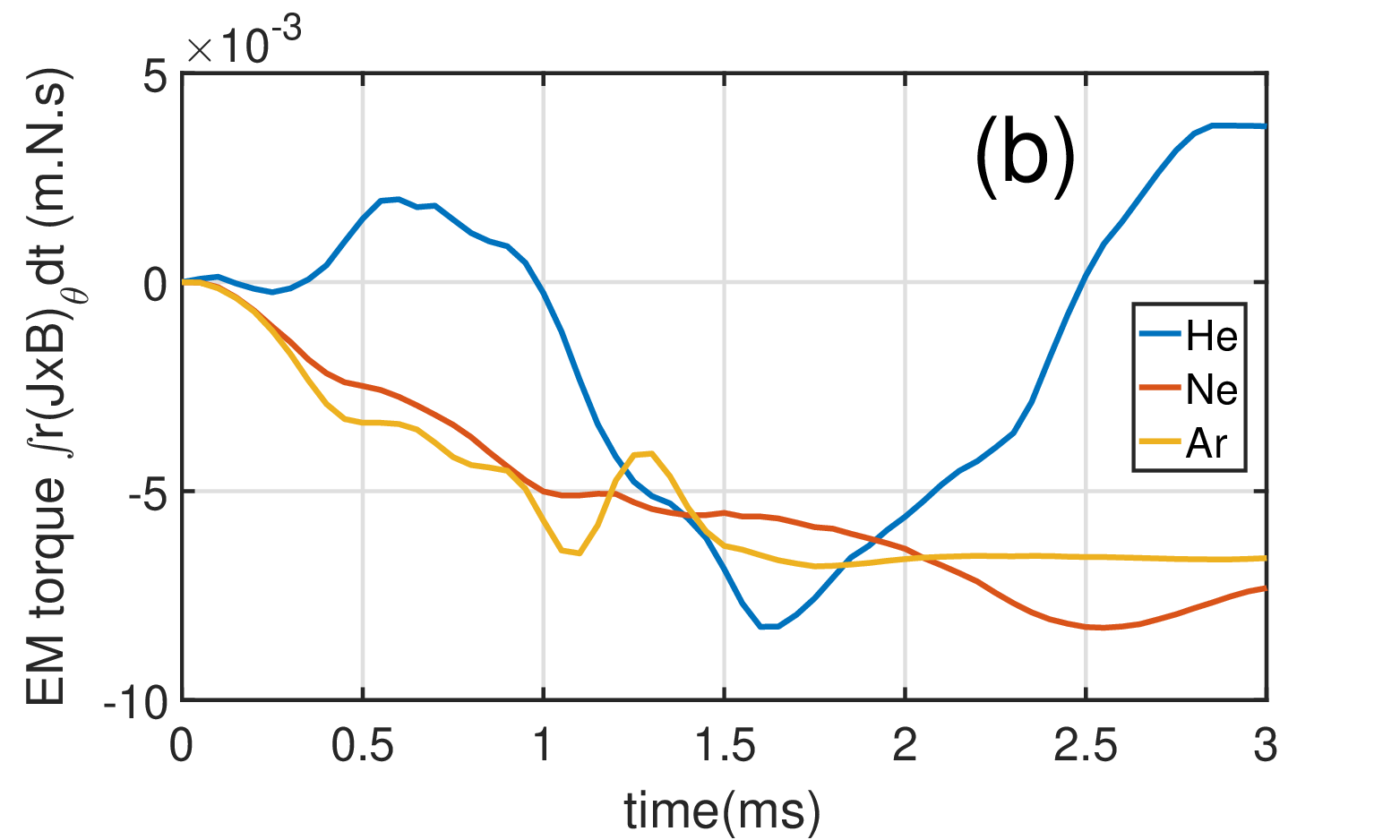}
		\end{center}
		\caption{(a) Impurity density peak poloidal angle in the injection poloidal plane $\theta_{imp,max} = \max(\int_{0}^{a}n_{imp}(r,\theta,\phi=0) dr/a)$, where $\theta=0$ is the outer mid-plane and $\theta=3\pi/2$ is the plasma bottom of impurity injection location, and (b) the time integral of the electromagnetic torque on $q=2$ surface $\int r(J\times B)_{\theta}dt$ for three impurity species as functions of time from 2D simulation cases.}
		\label{fig:2D-imp peak torque}
	\end{figure}

	We first consider the interaction between the impurity injection and the plasma in absence of any non-axisymmetric instability based on 2D simulations including only the axisymmetric Fourier mode with the toroidal mode number $n=0$. Three cases of simulations are set up with Helium, Neon, and Argon gas introduced separately to each case respectively. The static neutral impurity gas is deposited right at the beginning of simulation without any initial velocity, which then becomes ionized upon interaction with plasma. The inward penetration of impurity ions at the early stage are mainly through diffusion as a result of its large localized density gradient at the bottom plasma edge region, which is followed by the inward convection further along the radial direction. Only for the low-Z species, namely, the He impurity, its concentration prefers to drift to the LFS along the poloidal direction in the poloidal plane first ($t=0\sim 1ms$), then reverse towards the HFS like other high-Z impurity species, i.e. Ne and Ar (Fig. \ref{fig:2D-imp peak torque}a).
	Such an impurity poloidal drift direction can be understood from the poloidal component of electromagnetic (EM) torque, where the poloidal force component $\left( \vec{J}\times \vec{B} \right)_{\theta} = J_{\phi}\times B_r - J_r \times B_{\phi}$, which can be directly affected by the impurity injection itself. Here the cylindrical-like toroidal coordinate system $(r,\theta,\phi)$ is adopted. The He-induced torque drives the impurity to the positive poloidal direction during the first $t=0\sim 1ms$ period and then to the opposite direction afterwards. Other high-Z impurity induced torques remain in the same direction over time, which is consistent with the poloidal drift direction of their density concentrations (Fig. \ref{fig:2D-imp peak torque}b).
	
\begin{figure}[ht]
	\begin{center}
		\includegraphics[width=0.85\textwidth,height=0.35\textheight]{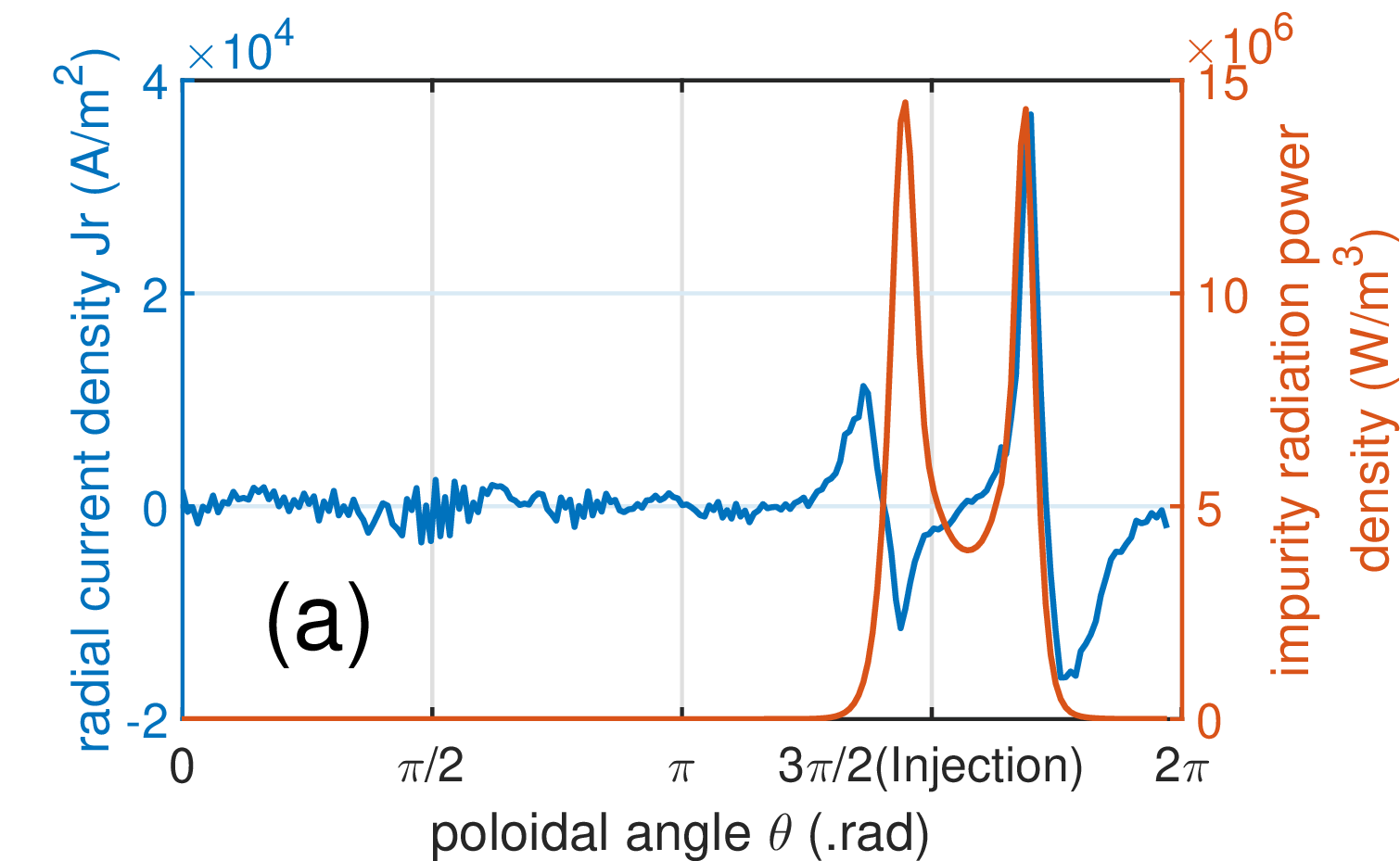}
		\includegraphics[width=0.85\textwidth,height=0.35\textheight]{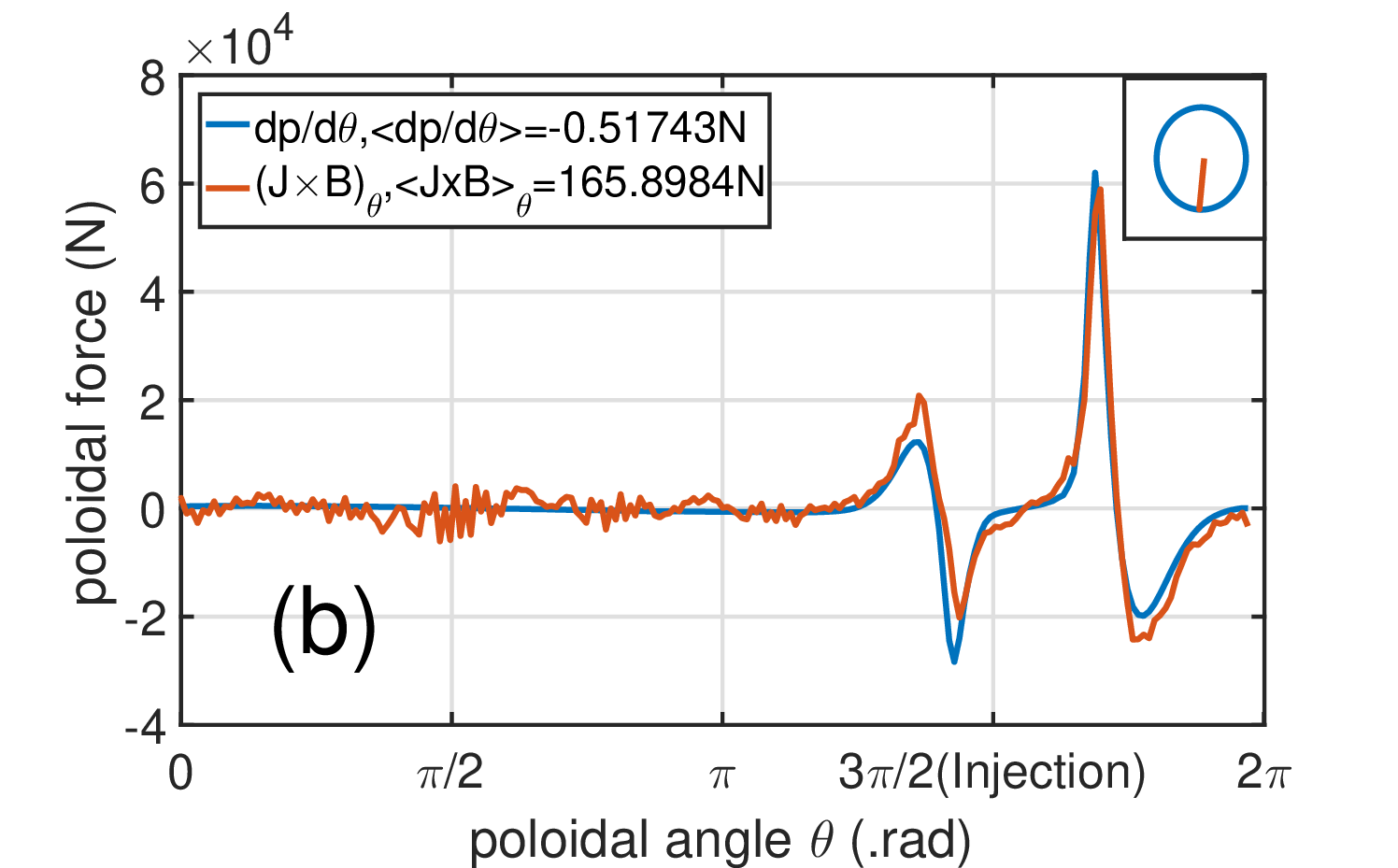}
	\end{center}
	\caption{(a) Impurity radiation power density (orange line), radial current density (blue line), (b) poloidal component of Lorentz force $(\vec{J}\times \vec{B})_{\theta}$ (orange line), and poloidal pressure gradient $dp/d\theta$ (blue line) on the $q=2$ surface at $t=0.5 ms$ for He impurity from 2D cases, $\theta=3\pi/2$ is the impurity injection location. The flux surface averaged pressure gradient $\left\langle dp/d\theta\right\rangle =0.51743N$, and the poloidal component of Lorentz force $\left\langle \vec{J}\times \vec{B}\right\rangle_{\theta}=165.8984N$.}
	\label{fig:2D-prad Jr force poloidal dis}
\end{figure}

	In particular, for all the impurity species the net or surface-averaged poloidal EM torque is dominated by the $J_r\times B_{\phi}$ component, which in turn primarily comes from the radial current density perturbation $J_r$.
	Impurity injection from the bottom edge leads to localized radiation cooling on each penetrated flux surface (Figs. \ref{fig:equilibrium}b and \ref{fig:2D-prad Jr force poloidal dis}a). The subsequent perturbation in the pressure and its gradient on the surface
	soon settle toward a new equilibrium, where the local balance between the pressure gradient and Lorentz forces along the poloidal direction is close to be exact (Fig. \ref{fig:2D-prad Jr force poloidal dis}b). The net residual imbalance between the surface-averaged force components, for example
	$\left\langle \vec{J}\times \vec{B}\right\rangle_{\theta}=165.8984N$ and $\left\langle dp/d\theta\right\rangle = -0.51743N$ at $t=0.5ms$, is relatively small in magnitude in comparison to the local peak values of the Lorentz and the pressure gradient forces that are the order of $10^4N$. Nonetheless, it is this net residual force imbalance, which is dominated by the Lorentz force $\left\langle \vec{J}\times \vec{B}\right\rangle_{\theta}$, that mainly contributes to the torque responsible for the driving of poloidal flow (Fig. \ref{fig:2D-prad Jr force poloidal dis}b).

\begin{figure}[ht]
		\begin{center}
			\includegraphics[width=0.45\textwidth,height=0.25\textheight]{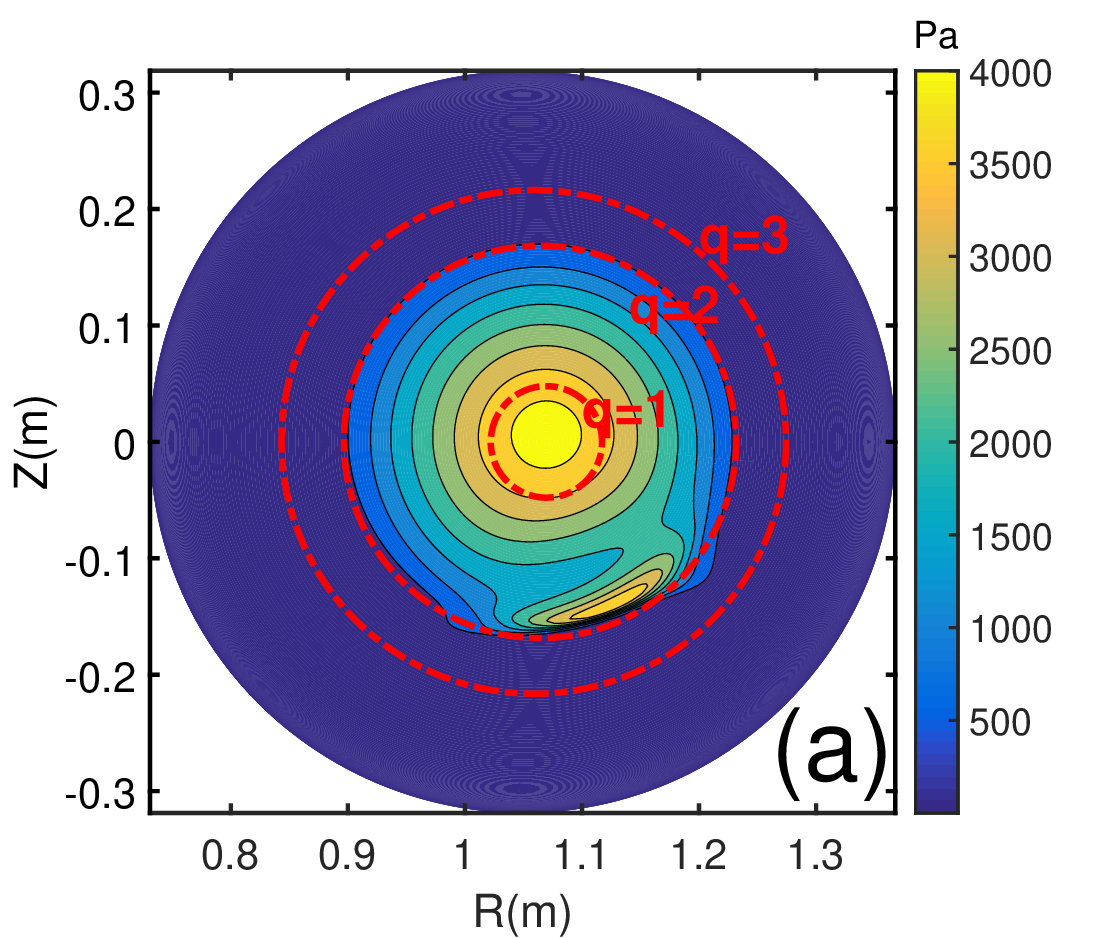}
			\includegraphics[width=0.45\textwidth,height=0.25\textheight]{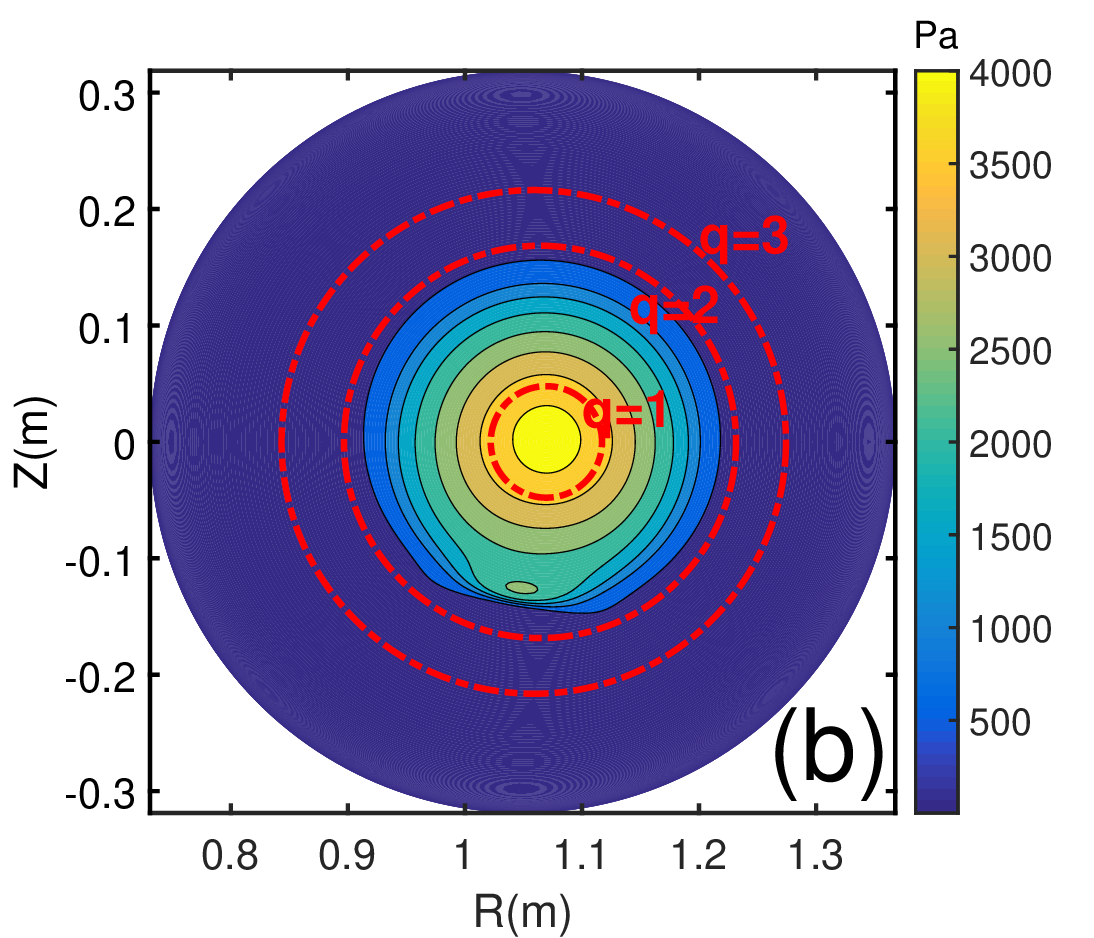}
			\includegraphics[width=0.45\textwidth,height=0.25\textheight]{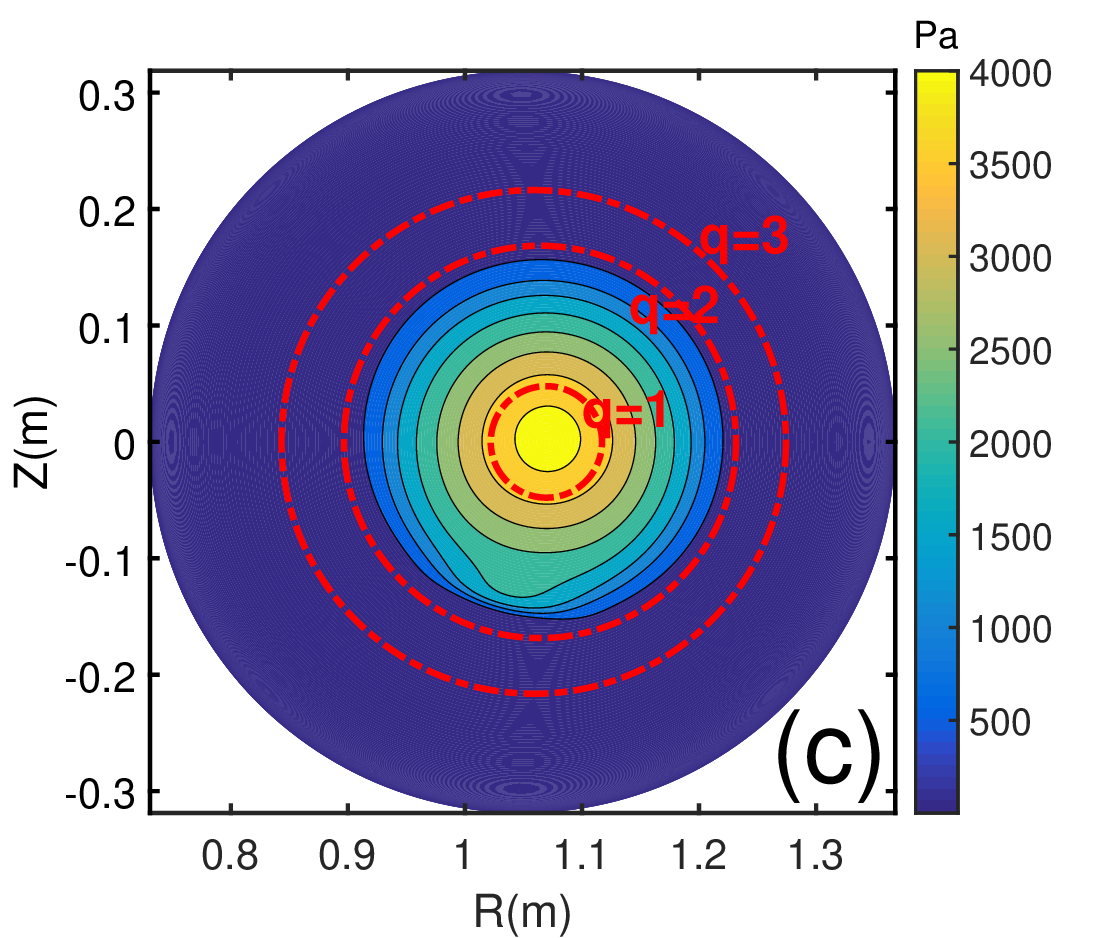}
		\end{center}
		\caption{Pressure contours at $t=0.5ms$ for (a) He, (b) Ne, and (c) Ar injections from 2D simulation cases, and the equilibrium $q=1,2,3$ surfaces are denoted as red dashed circles for references.}
		\label{fig:2D-pres contour}
	\end{figure}
\begin{figure}[ht]
		\begin{center}
			\includegraphics[width=0.85\textwidth,height=0.35\textheight]{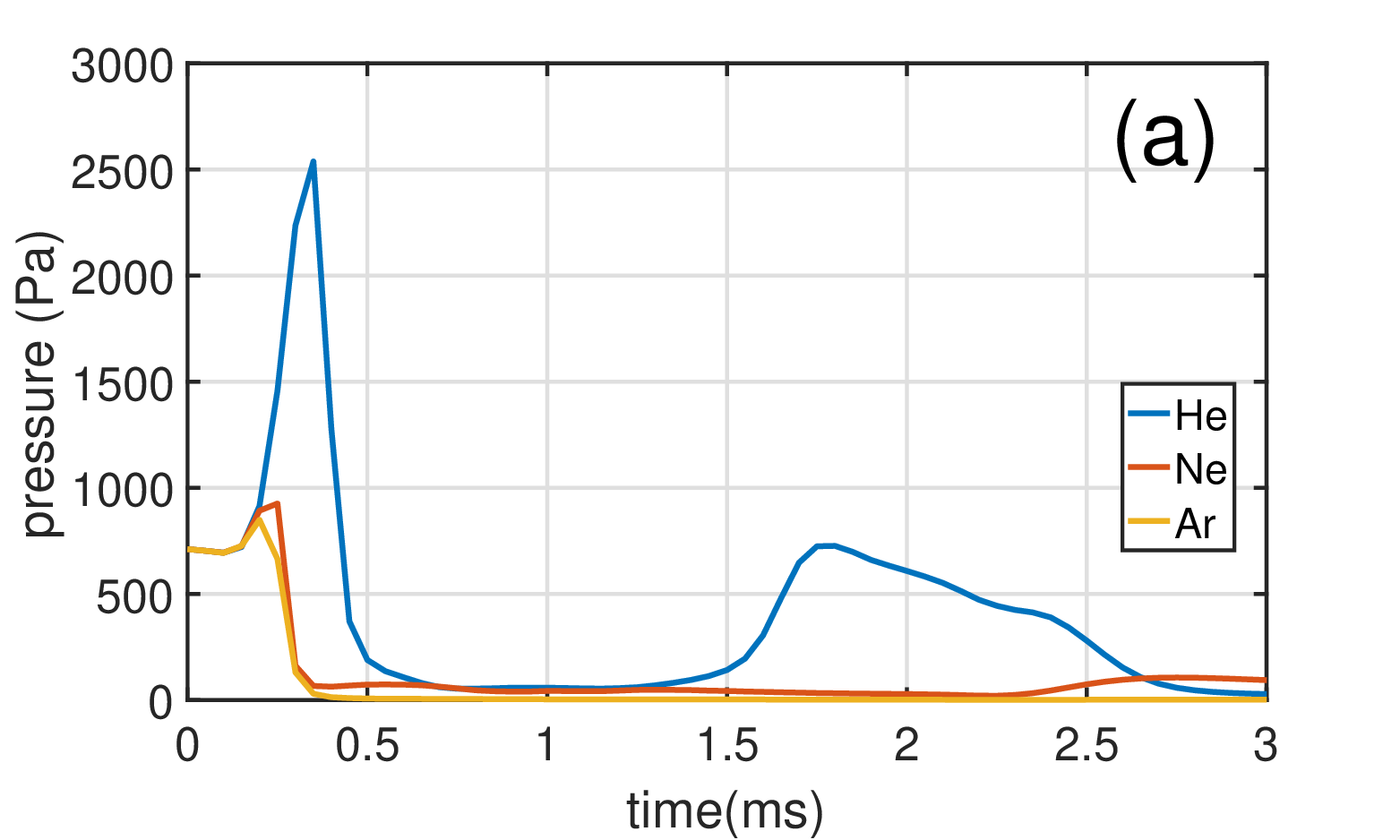}
			\includegraphics[width=0.85\textwidth,height=0.35\textheight]{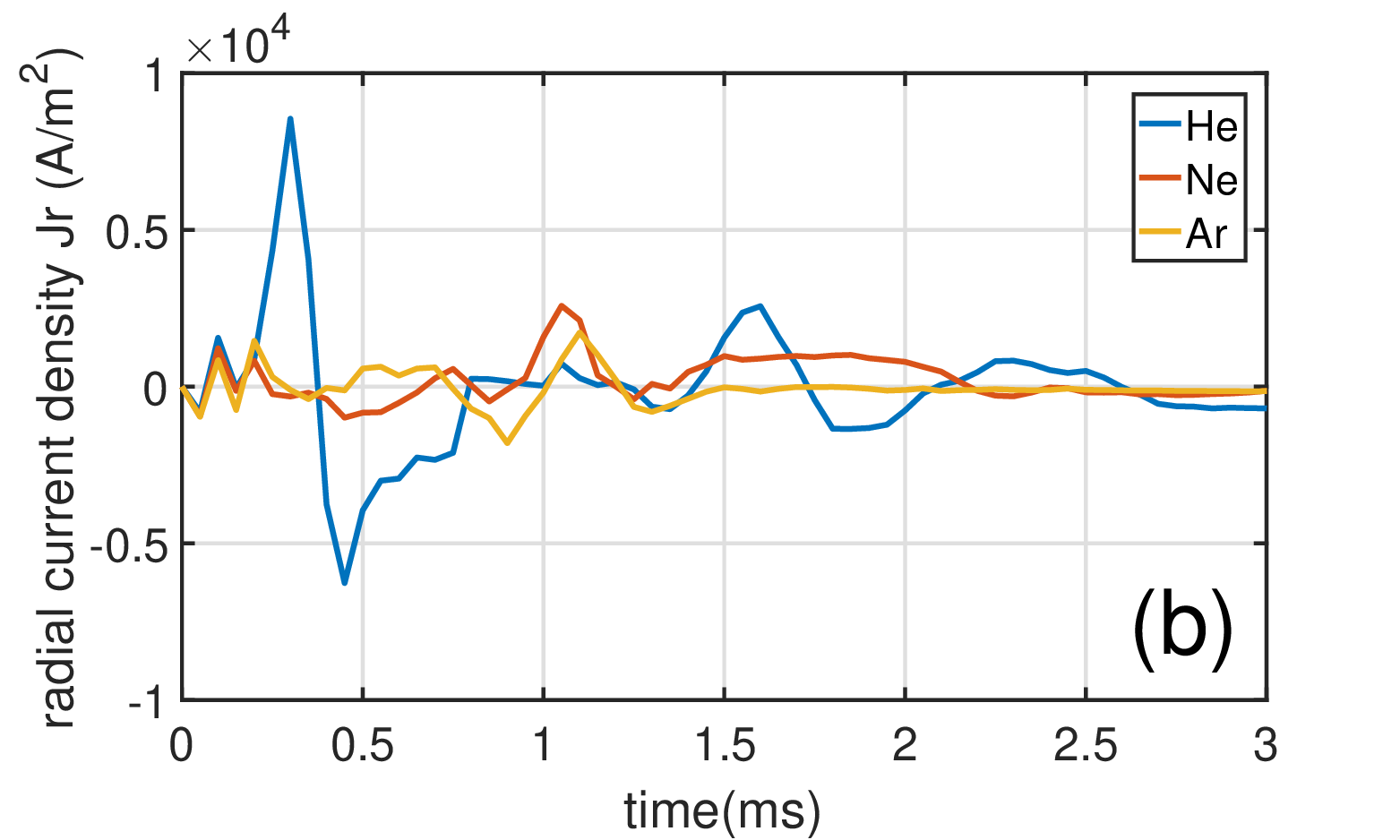}
		\end{center}
		\caption{(a) Local pressure $P$ and (b) radial current density $J_r$ at the bottom of the $q=2$ surface (i.e. $\theta=3\pi /2$) for three impurity species as functions of time from 2D simulation cases.}
		\label{fig:2D-pres Jr}
	\end{figure}

	Further inspection finds that the major difference in the poloidal torque balance above between low-Z and high-Z impurity species may come from the formation of a local high-$\beta$ plasmoid at the He injection location before the start of the density peak drift along the poloidal direction (Figs. \ref{fig:2D-pres contour} and \ref{fig:2D-pres Jr}a). This local enhanced pressure is mainly a consequence of the density perturbation due to the stronger He impurity ionization in the temperature range of the plasma edge region. Through the local equilibrium force balance, i.e. $J_r \approx -\partial_{\theta} p/B_{\phi}$, this high-$\beta$ plasmoid results in a much larger perturbation to the poloidal pressure gradient and the radial component of the current density in comparisons to the impurity injection processes involving other higher-Z species.
	The He impurity poloidal flow reversal is directly caused by the direction change of the radial current density perturbation $J_r$ over time, since the toroidal magnetic field $B_{\phi}$ remains in the same direction over the same time frame (Fig. \ref{fig:2D-pres Jr}b).

    \section{Impurity injection induced poloidal flow and the magnetic island rotation in the poloidal plane}
	\label{Sec: 3D cases}
	
	3D simulations including Fourier components with toroidal mode numbers $n=0-5$ are considered here, and all other setups remain same as the 2D simulations. The $n=1$ mode dominates the entire impurity penetration process, and the poloidal mode number $m=2$ tearing mode is induced by the impurity radiation and becomes the dominant MHD instability rapidly after the impurity injection \cite{Zeng_2023_tm,Zeng_2023}. The main behavior of impurity density penetration is similar to the 2D cases, even though some differences appear in the presence of magnetic islands.

	\subsection{Correlation between the impurity distribution, the poloidal flow, and the island rotation}
\begin{figure}[ht]
		\begin{center}
			\includegraphics[width=0.45\textwidth,height=0.25\textheight]{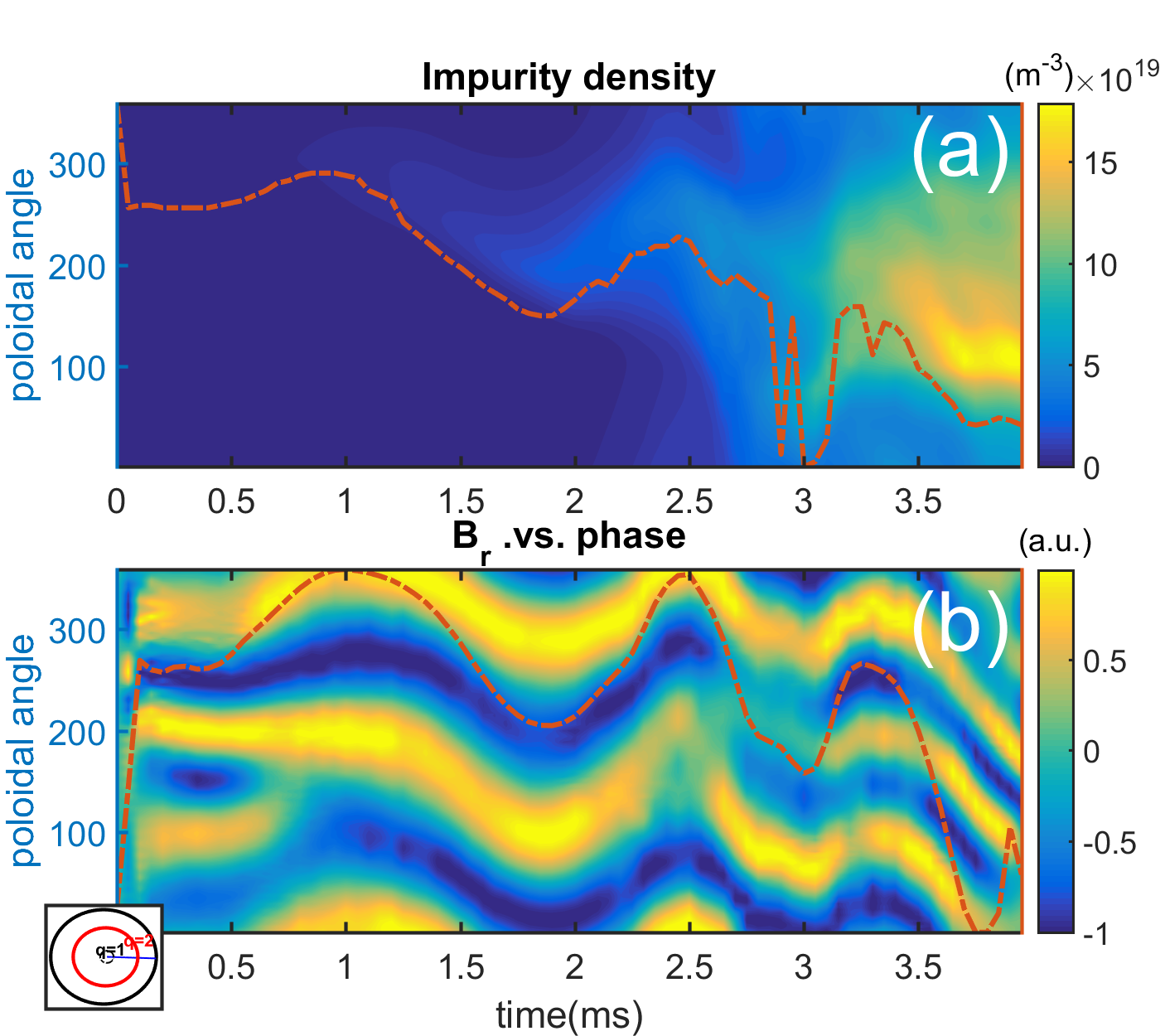}
			\includegraphics[width=0.45\textwidth,height=0.25\textheight]{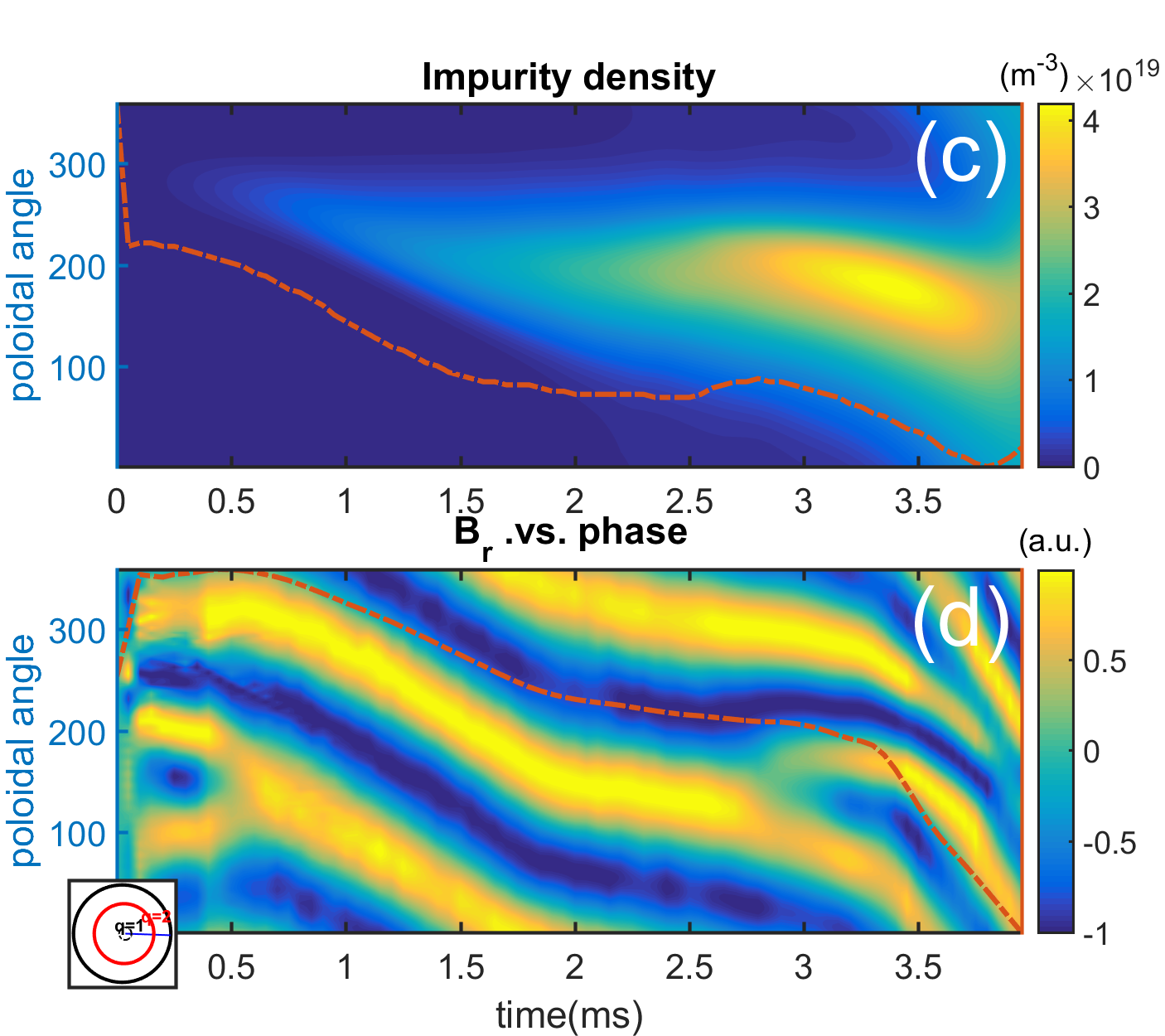}
			\includegraphics[width=0.45\textwidth,height=0.25\textheight]{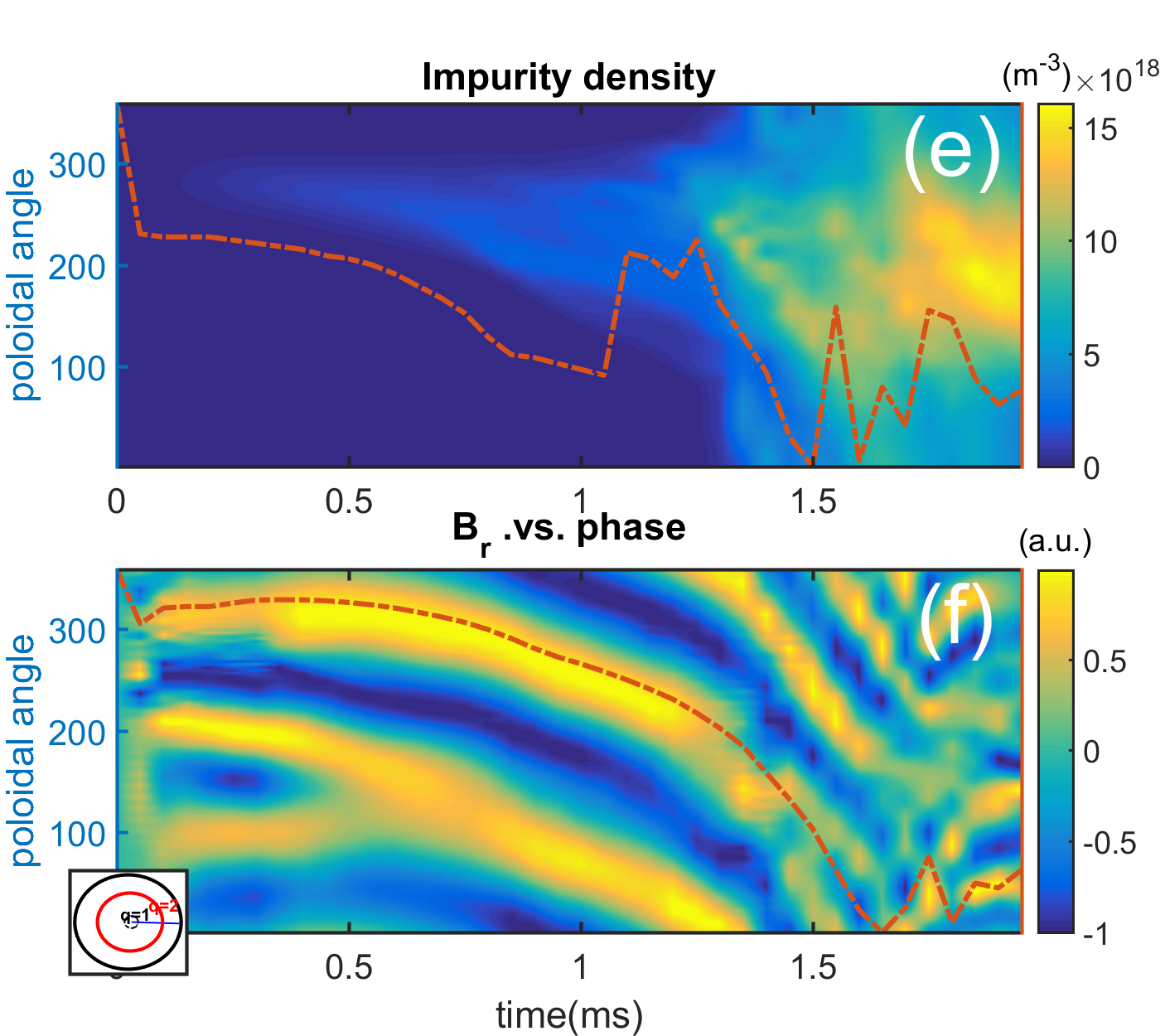}
		\end{center}
		\caption{Poloidal distributions of (a), (c), (e) impurity density $n_{imp}(\theta) = \int_{0}^{a}n_{imp}(r,\theta,\phi=0) dr/a$, and (b), (d), (f) the normalized $n=1$ component of radial magnetic field $B_r$ as functions of time for He (a,b), Ne(c,d) and Ar(e,f) injection, respectively from 3D simulation cases. The orange dashed-lines denote the poloidal angles of the impurity density peak and the phase of the $B_r$. The unit of the poloidal angle is degree, $\theta=0$ is the outer mid-plane and $\theta=270$ is the plasma bottom of impurity injection location.}
		\label{fig:3D-mode imp pol dis}
	\end{figure}

	As shown in Fig. \ref{fig:3D-mode imp pol dis}, only low-Z He impurity density peak tends to drift towards the LFS first and then reverse to the HFS later, whereas both Ne and Ar impurity density peaks prefer to drift towards the HFS only, which is similar to the 2D simulations. However, the He impurity density drift reverses several times along the poloidal direction over the same time period in the 3D simulation. The phase of the dominant $n=1$ mode reverses along with the He impurity density poloidal drift reversal in the poloidal direction, whereas no mode phase reversal has been observed in the Ne or Ar case, which agrees with the experimental observations \cite{Li_2020}.

\begin{figure}[ht]
		\begin{center}
			\includegraphics[width=0.7\textwidth,height=0.25\textheight]{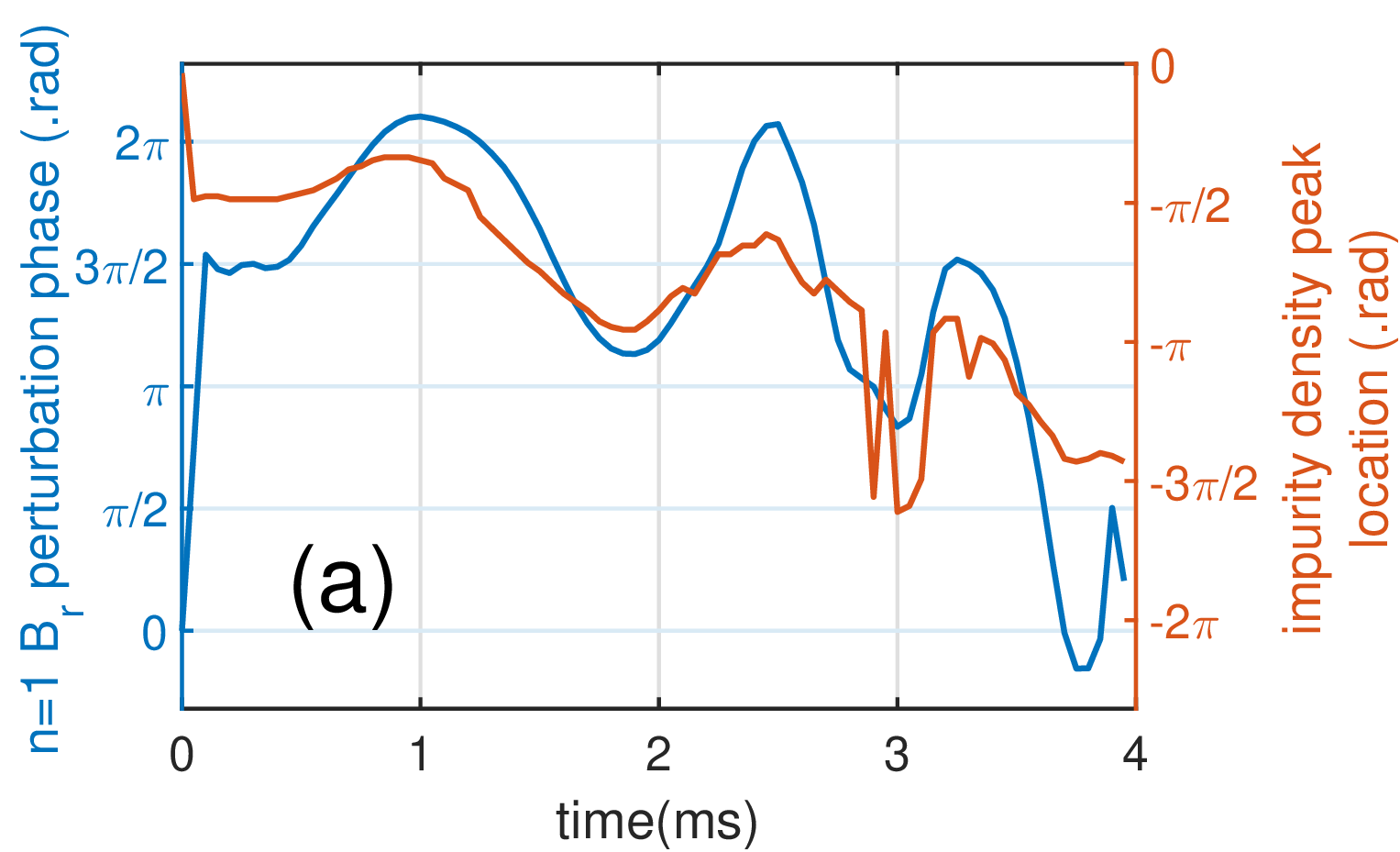}
			\includegraphics[width=0.7\textwidth,height=0.25\textheight]{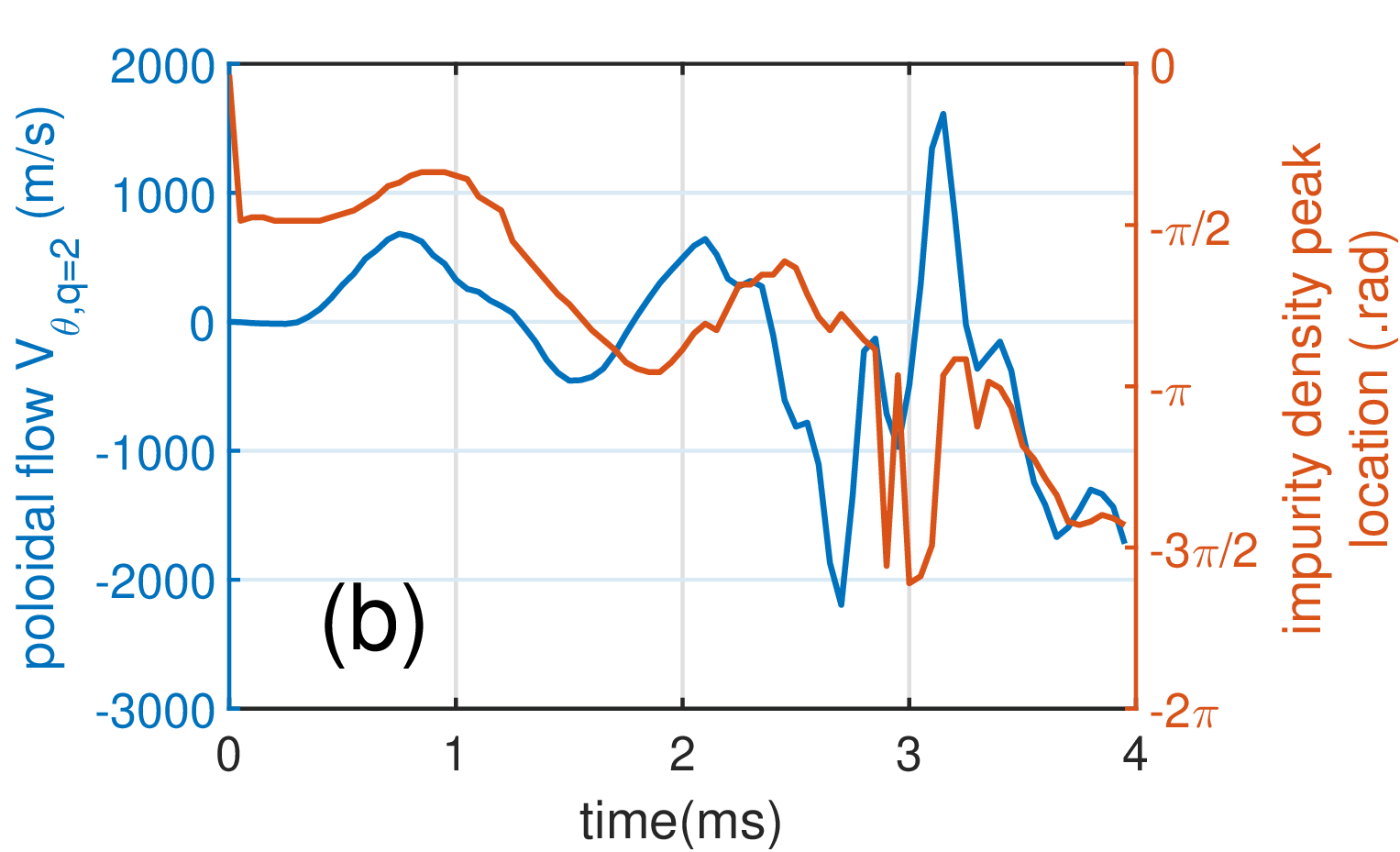}
			\includegraphics[width=0.7\textwidth,height=0.25\textheight]{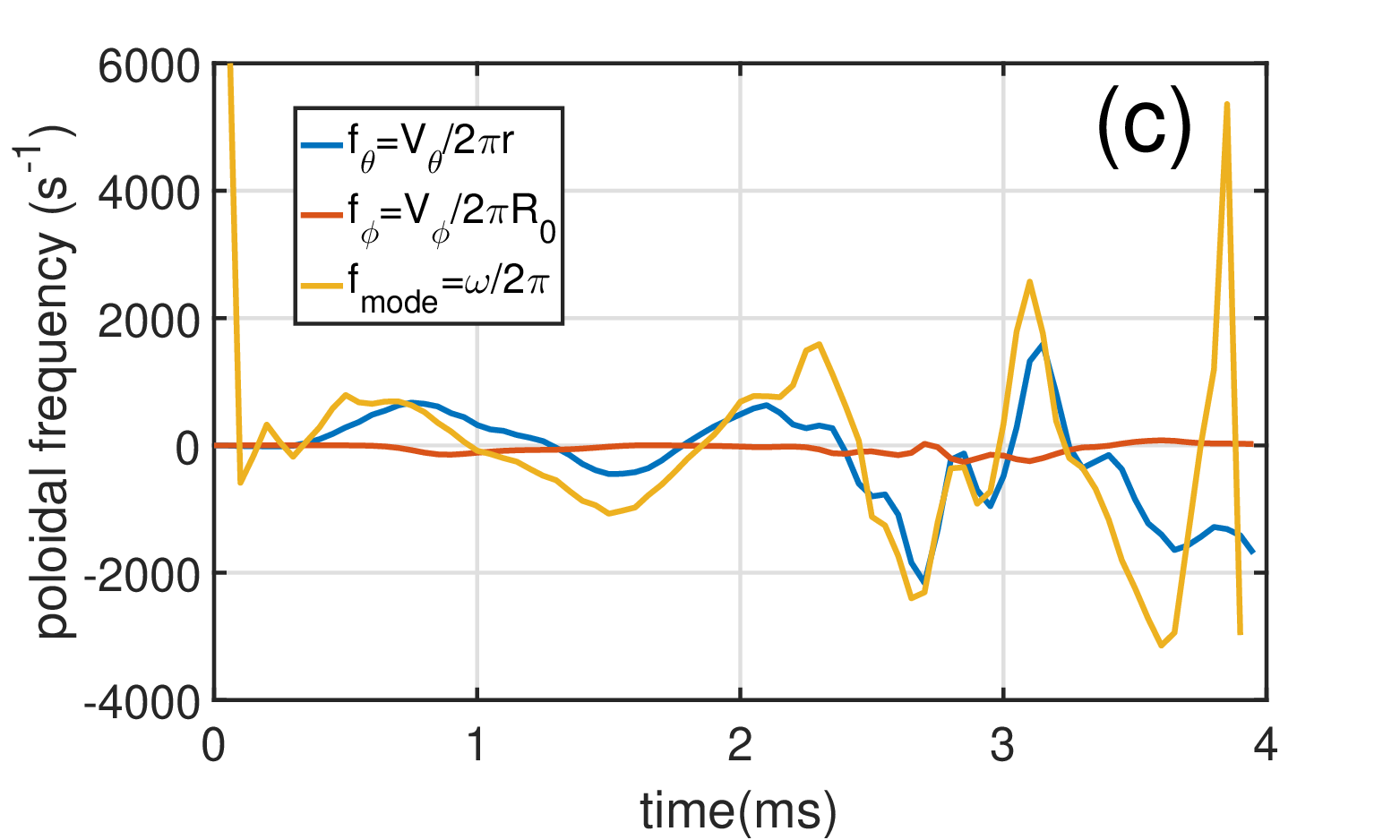}
		\end{center}
		\caption{(a) Impurity density peak poloidal angle (orange line) and the phase of the $n=1$ component of radial magnetic field $B_r$ (blue line), (b) impurity density peak poloidal angle (orange line) and the poloidal flow $V_{\theta}$ on the $q=2$ surface (blue line), (c) frequencies of the poloidal flow (blue line), toroidal flow (orange line), and the $n=1$ component of $B_r$ (yellow line) as functions of time for the He impurity from the 3D simulation case.}
		\label{fig:3D-correlations}
	\end{figure}

	The poloidal phase variation of the $n=1$ mode in time shows close correlation with that of the impurity density peak in the poloidal plane (Fig. \ref{fig:3D-correlations}a). And the perturbed poloidal flow, which is peaked around the $q=2$ surface, also tracks the time evolution of the impurity density peak poloidal angle (Fig. \ref{fig:3D-correlations}b). Correspondingly, the frequencies of the poloidal flow and the MHD mode agree well most of time as shown in Fig. \ref{fig:3D-correlations}(c), which is consistent with their both correlations to the impurity density distribution. It is well known that impurity gas injection is shallow and the gas cold front usually stops along the $q=2$ surface location \cite{Reux_2010}, probably because of this the poloidal flow peaked around the $q=2$ surface, which is synchronized with the induced dominant $m=2/n=1$ tearing mode.
	The toroidal flow is negligible in comparison to the island rotation (Fig. \ref{fig:3D-correlations}c), and this poloidal flow-island dynamic discussed here is almost irrelevant to the toroidal evolution.
	Beyond the above correlations, the understanding of the causal relations among the impurity density, the poloidal flow, and the magnetic island rotation frequency may require further torque balance analysis next.

	\subsection{The electromagnetic torque produced by the impurity radiation}
	
\begin{figure}[ht]
		\begin{center}
			\includegraphics[width=0.7\textwidth,height=0.25\textheight]{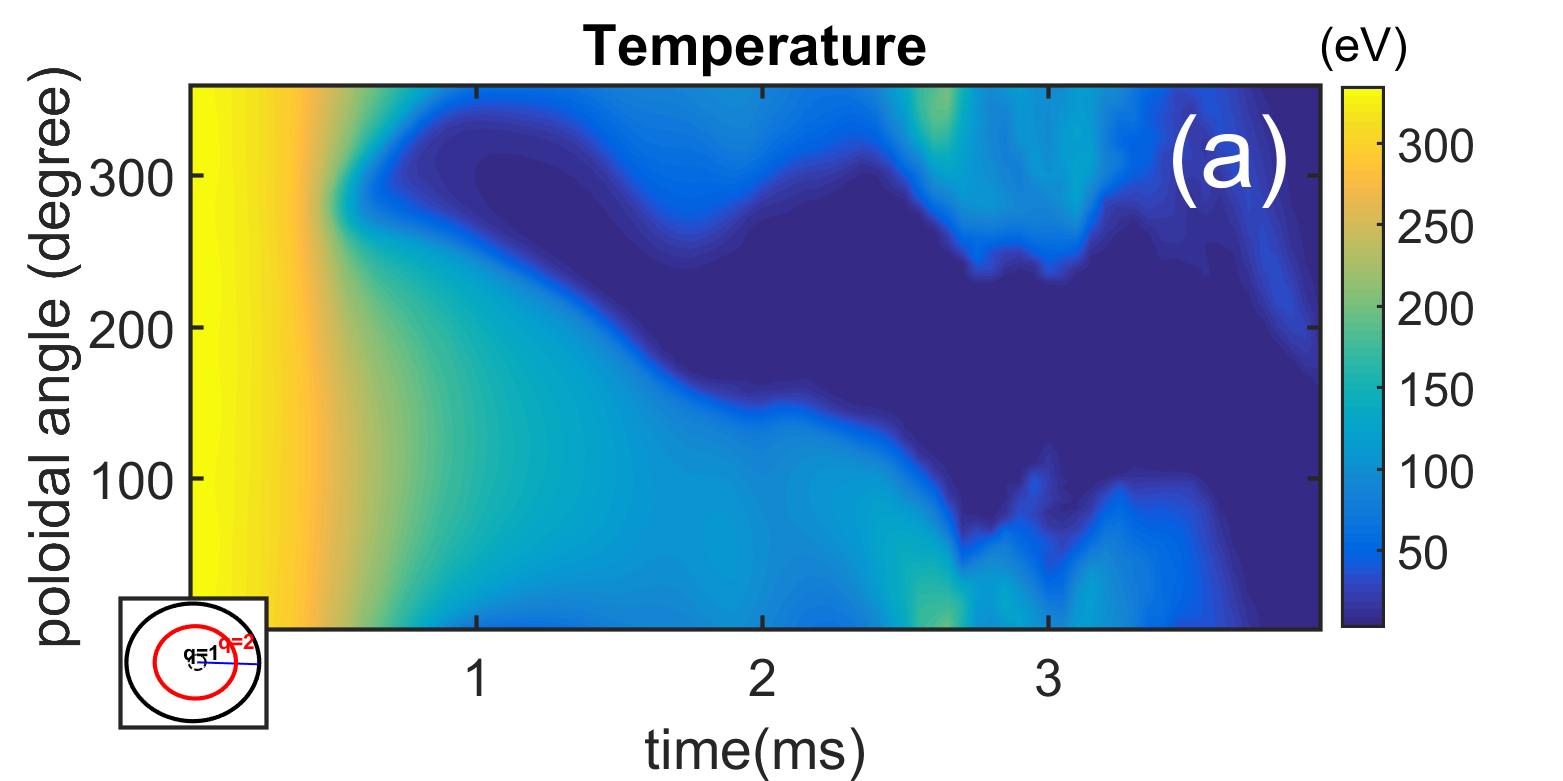}
			\includegraphics[width=0.7\textwidth,height=0.25\textheight]{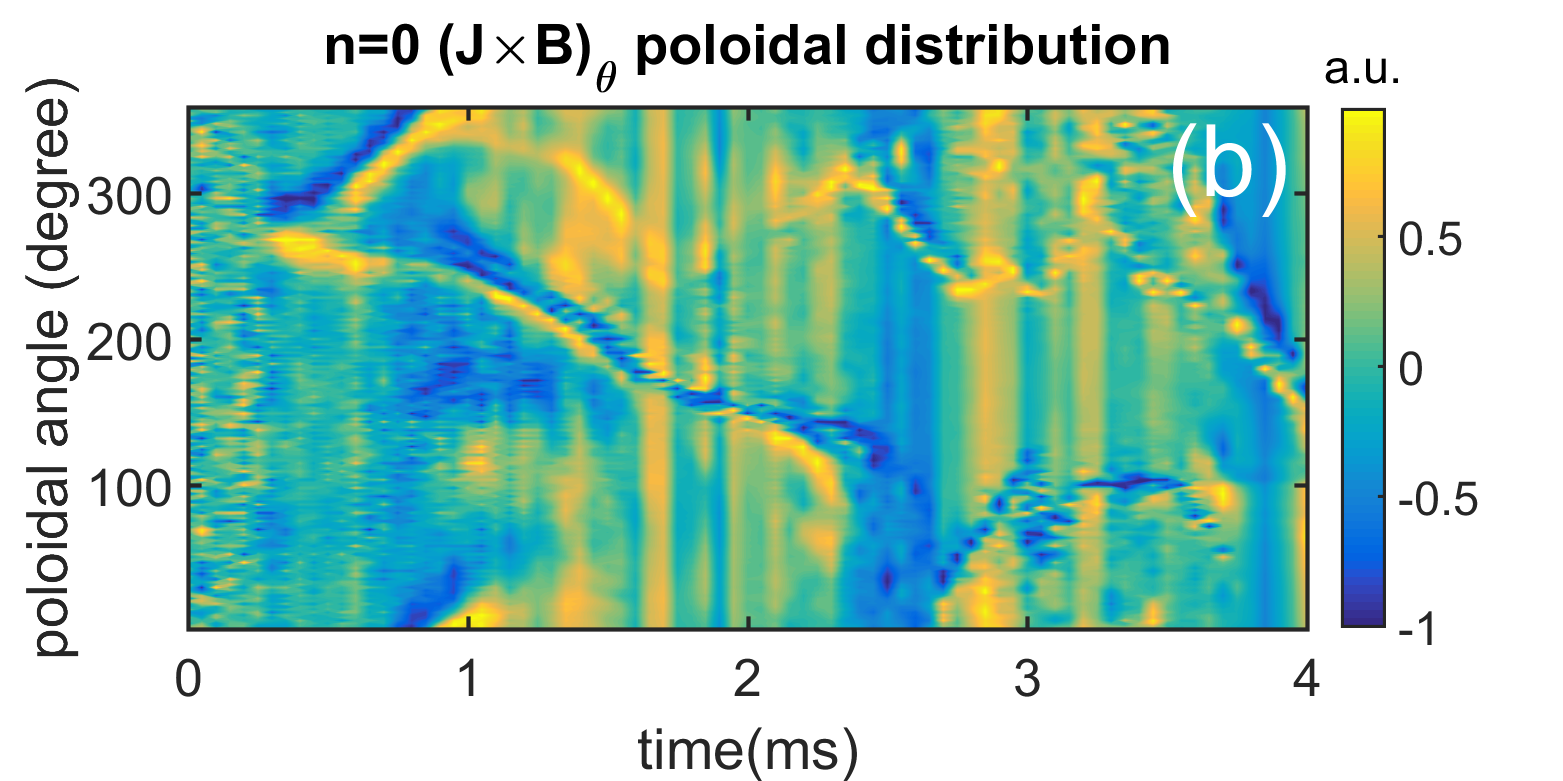}
			\includegraphics[width=0.7\textwidth,height=0.25\textheight]{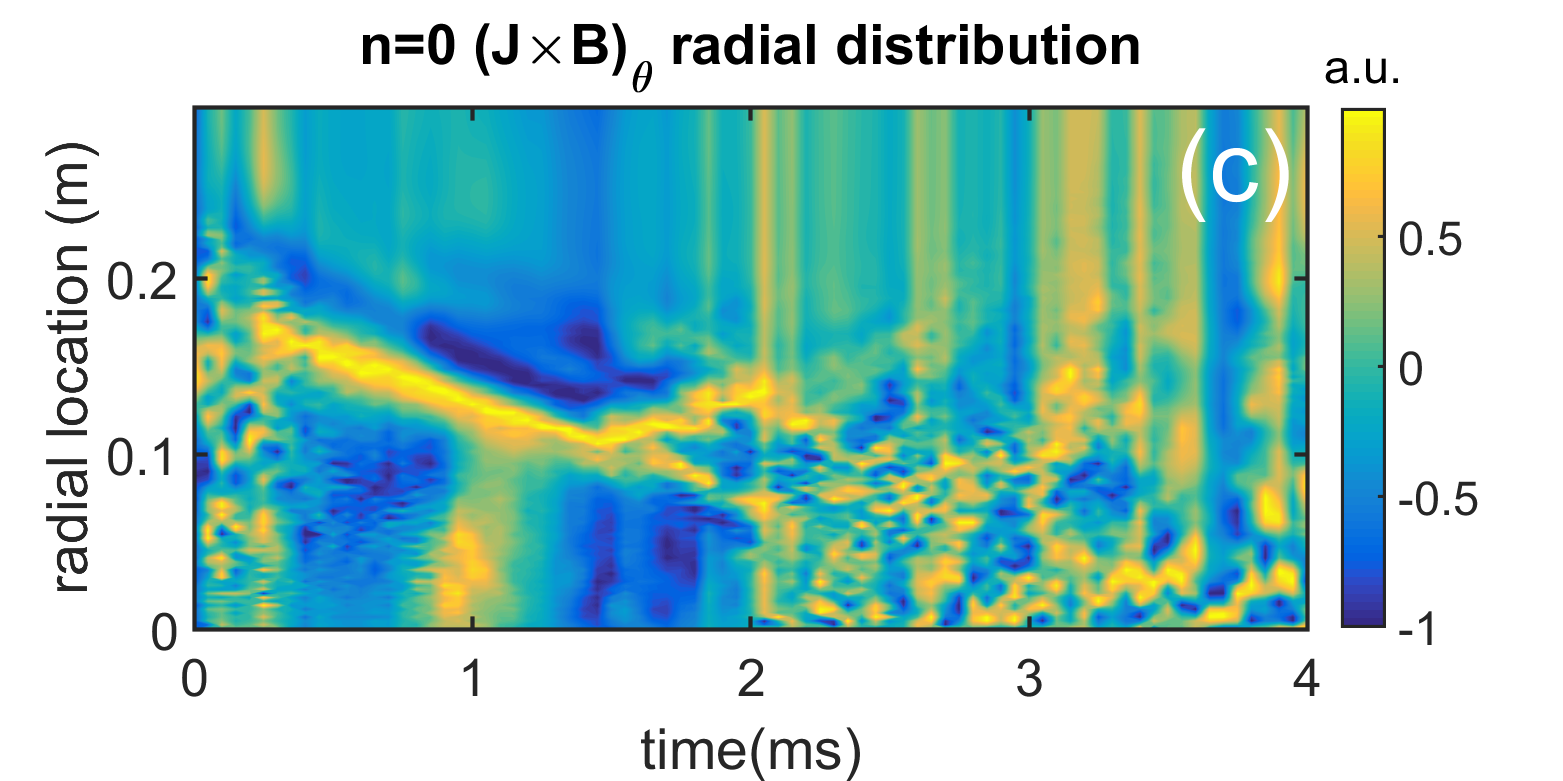}
		\end{center}
		\caption{Poloidal distributions of (a) the electron temperature and (b) the normalized $n=0$ component of poloidal Lorentz force $(\vec{J} \times \vec{B})_{\theta}$ on the $q=2$ surface, and the radial distribution of (c) the surface-averaged normalized $n=0$ component of poloidal Lorentz force $\left\langle \vec{J} \times \vec{B}\right\rangle_{\theta} = \int_0^{2\pi} (\vec{J} \times \vec{B})_{\theta} d\theta/2\pi$ as functions of time for the He impurity from 3D simulation case.}
		\label{fig:torque spatial dis}
	\end{figure}

	The impurity injection leads to the density and temperature perturbation directly, modifies the local equilibrium and thus induces current density and magnetic field perturbations. The poloidal distribution of the $n=0$ component of the EM force produced by such an impurity injection is shown in Fig. \ref{fig:torque spatial dis}, where the variation of the local enhanced EM force on the surface is consistent with locations of the localized impurity radiation cooling and the consequent sharp temperature gradient on the surface (Figs. \ref{fig:torque spatial dis}a-b). In addition, the radial peak location of the surface-averaged EM force moves inward along with the impurity penetration (Fig. \ref{fig:torque spatial dis}c), and it is well aligned with the impurity cold front during the early $t=0-2ms$ period, until it becomes radially scattered afterwards when the TQ fully develops with complete stochastisation of magnetic field lines. The spatial distribution of the $n=0$ component of EM force is strongly correlated with the impurity radiation cooling in both the radial and poloidal directions, suggesting that this EM force component is the main driver for the poloidal flow and the island rotation.

\begin{figure}[ht]
		\begin{center}
			\includegraphics[width=0.85\textwidth,height=0.35\textheight]{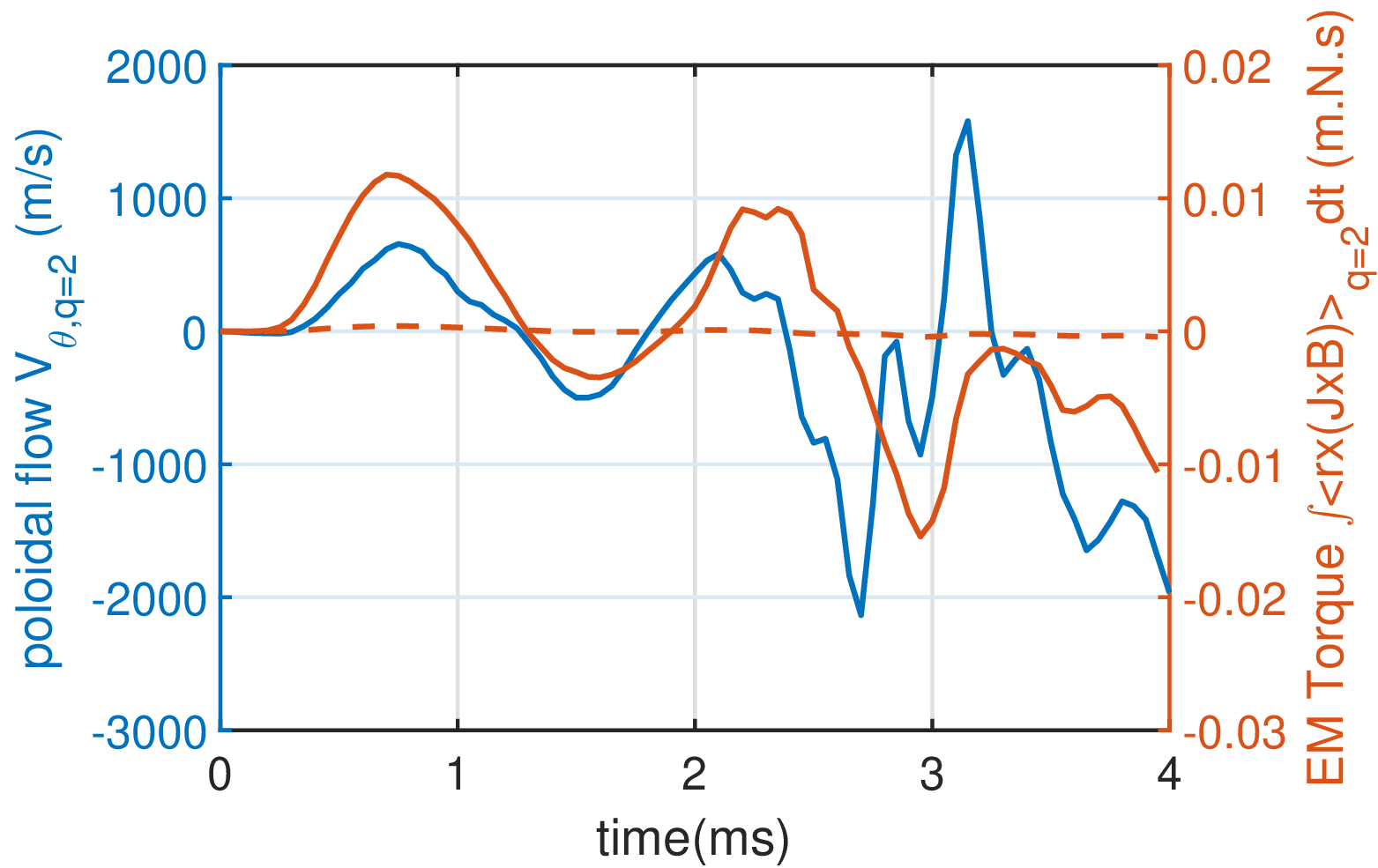}
		\end{center}
		\caption{The poloidal flow $V_{\theta}$ on the $q=2$ surface (blue solid line) and the time integral of the $n=0$ component of electromagnetic torque averaged over the $2/1$ island region $ \int \left\langle r( \vec{J}\times \vec{B})_{\theta} \right\rangle_{q=2} dt  = \int_0^{\tau} \left( \int_0^{2\pi} \left( \int_{r_s-w/2}^{r_s+w/2}  r(\vec{J}\times \vec{B})_{\theta} dr/w \right)  d\theta/2\pi \right) dt $ (orange solid line), and the poloidal pressure gradient averaged over the island region $ \int \left\langle r(dp/d\theta)\right\rangle_{q=2} dt = \int_0^{\tau} \left( \int_0^{2\pi} \left( \int_{r_s-w/2}^{r_s+w/2}  r(dp/d\theta)_{\theta} dr/w \right)  d\theta/2\pi \right) dt$ (orange dashed-line) as functions of time for the He impurity from the 3D simulation case.}
		\label{fig:3D-torque vtheta}
	\end{figure}

	The time integral of the island region averaged EM torque agrees with the overall variation of the poloidal flow in time (Fig. \ref{fig:3D-torque vtheta}).
	In the meantime, the same averaged poloidal pressure gradient is shown negligible in comparison to the EM force. The poloidal flow perturbation and its poloidal gradient are zero in the initial static equilibrium, and the torque comes entirely from the perturbation.

\begin{figure}[ht]
		\begin{center}
			\includegraphics[width=0.85\textwidth,height=0.35\textheight]{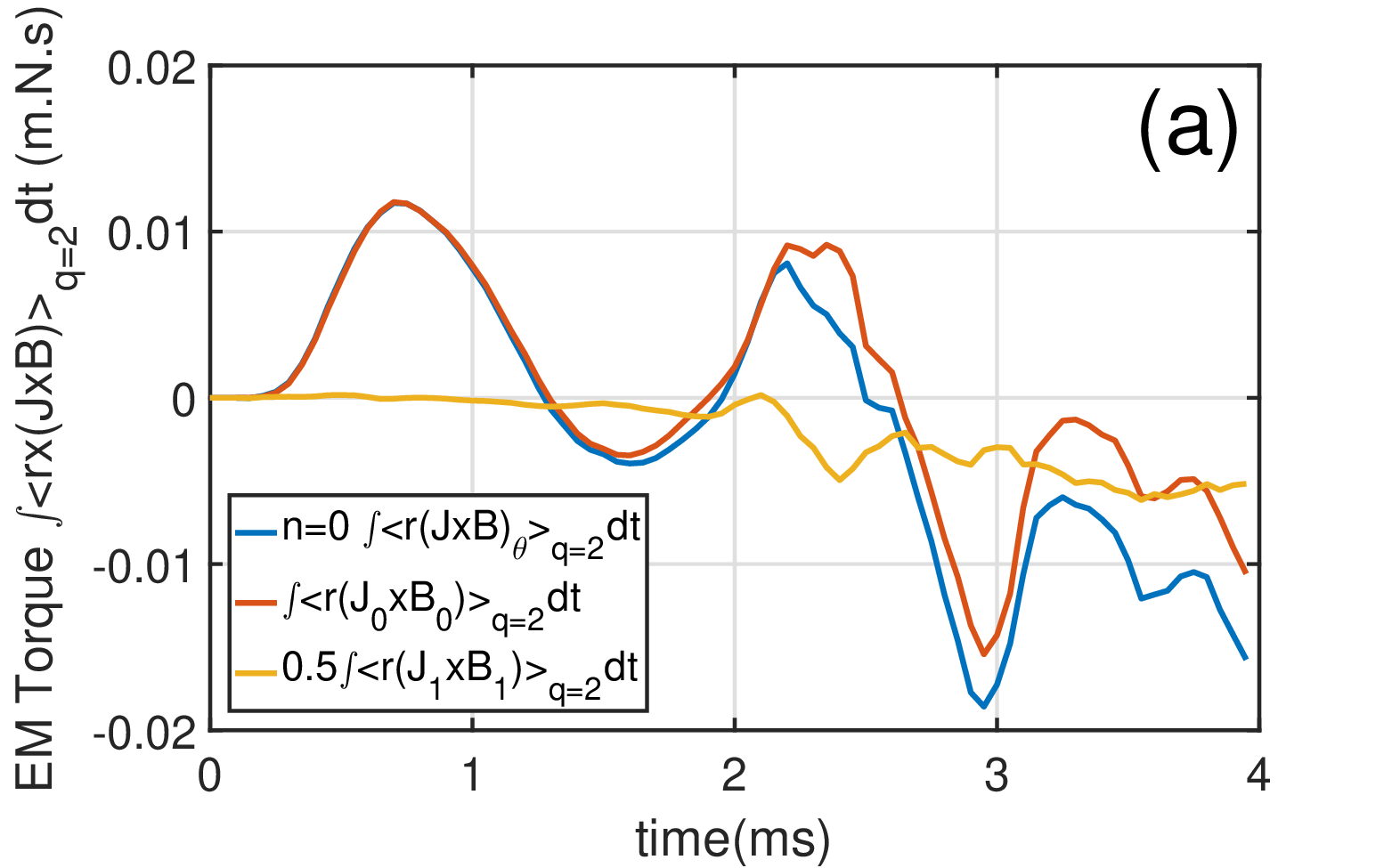}
			\includegraphics[width=0.85\textwidth,height=0.35\textheight]{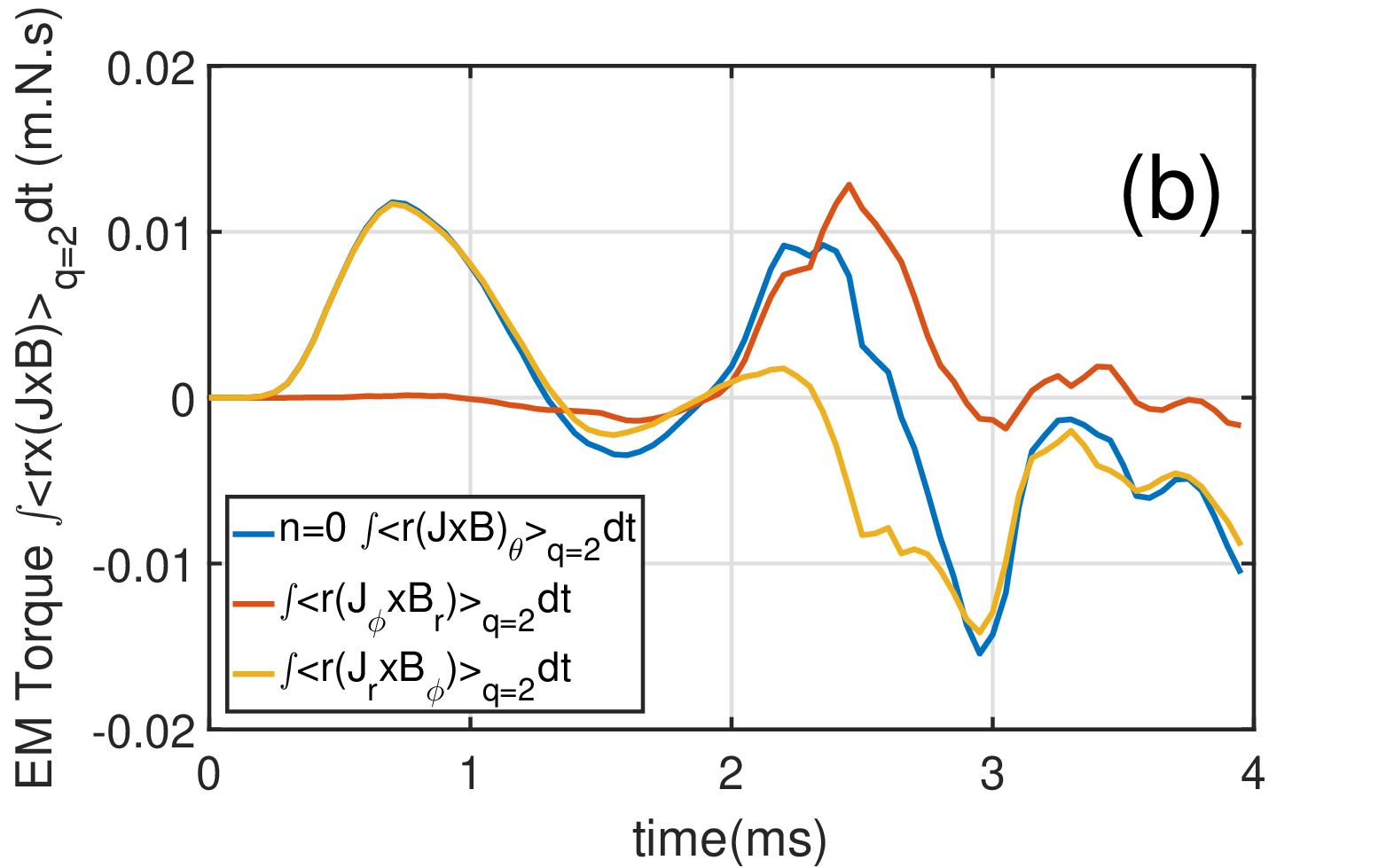}
		\end{center}
		\caption{(a) The time integral of the $n=0$ component of electromagnetic torque averaged over the $2/1$ island region $\int \left\langle r(\vec{J}\times \vec{B})_{\theta}\right\rangle_{q=2}dt$ (blue line) $= \int \left\langle  r(\vec{J_0}\times \vec{B_0}) \right\rangle_{q=2} dt$ (red line) $+ 0.5 \int \left\langle r(\vec{J_1}\times \vec{B_1}) \right\rangle_{q=2} dt$ (yellow line), where $J_0, J_1$ and $B_0, B_1$ are the $n=0$ and $n=1$ components of current and magnetic field, respectively, and (b) the time integral of the $n=0$ component of electromagnetic torque averaged over the island region $\int\left\langle  r(\vec{J}\times \vec{B})_{\theta} \right\rangle_{q=2}dt$ (blue line) $= \int \left\langle  r(J_{\phi}\times B_r) \right\rangle_{q=2} dt$ (red line) $+\int \left\langle r(J_r\times B_{\phi}) \right\rangle_{q=2}dt$ (yellow line) as functions of time for the He impurity from the 3D simulation case.}
		\label{fig:3D-torque analysis}
	\end{figure}

	A detailed EM force analysis is shown in Fig. \ref{fig:3D-torque analysis}, where $( \vec{J}\times \vec{B})_{\theta,n=0} \simeq (\vec{J_0}\times \vec{B_0})_{\theta} +(\vec{J_1}\times \vec{B_1})_{\theta}$, i.e. the $n=0$ component of the EM force mainly comes from the $n=0$ component of current density perturbation, which corresponds to the modification of the equilibrium profiles by the impurity radiation, and the nonlinear beating from the $n=1$ components after $t=2ms$ through the island saturation (Fig. \ref{fig:3D-torque analysis}a).
	On the other hand, the poloidal force $(\vec{J}\times \vec{B})_{\theta} = J_{\phi} \times B_r - J_r\times B_{\phi}$, and the current perturbation dominates the force variation, because the direction of $B_{\phi}$ can not change and the poloidal structure of the perturbed $B_r$ almost remain the same. The $J_r \times B_{\phi}$ part dominates the force during the early phase $t=0-2ms$, and the $J_{\phi} \times B_r$ part becomes comparable only after $t=2ms$ (Fig. \ref{fig:3D-torque analysis}b).
	
\begin{figure}[ht]
		\begin{center}
			\includegraphics[width=0.85\textwidth,height=0.35\textheight]{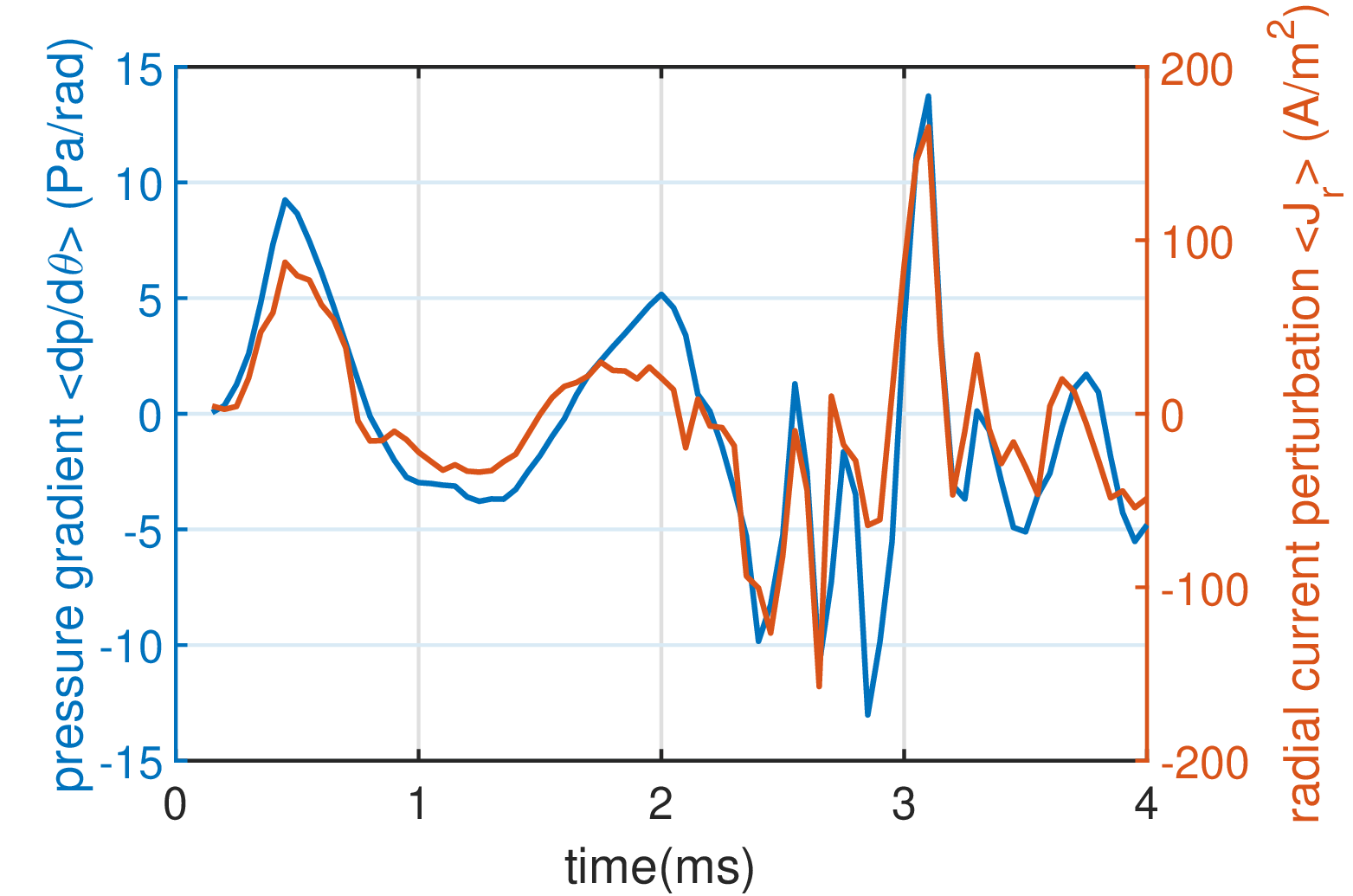}
		\end{center}
		\caption{The island region averaged poloidal pressure gradient $\left\langle dp/d\theta\right\rangle = \int_0^{2\pi}\left( \int_{r_s-w/2}^{r_s+w/2} (dp/d\theta) dr/w \right) d\theta/2\pi $ (blue line) and radial current density $\left\langle J_r\right\rangle =\int_0^{2\pi}\left( \int_{r_s-w/2}^{r_s+w/2} J_r dr/w \right) d\theta/2\pi $ (orange line) as functions of time for the He impurity from the 3D simulation case.}
		\label{fig:3D-Jr source}
	\end{figure}

\begin{figure}[ht]
		\begin{center}
			\includegraphics[width=0.85\textwidth,height=0.35\textheight]{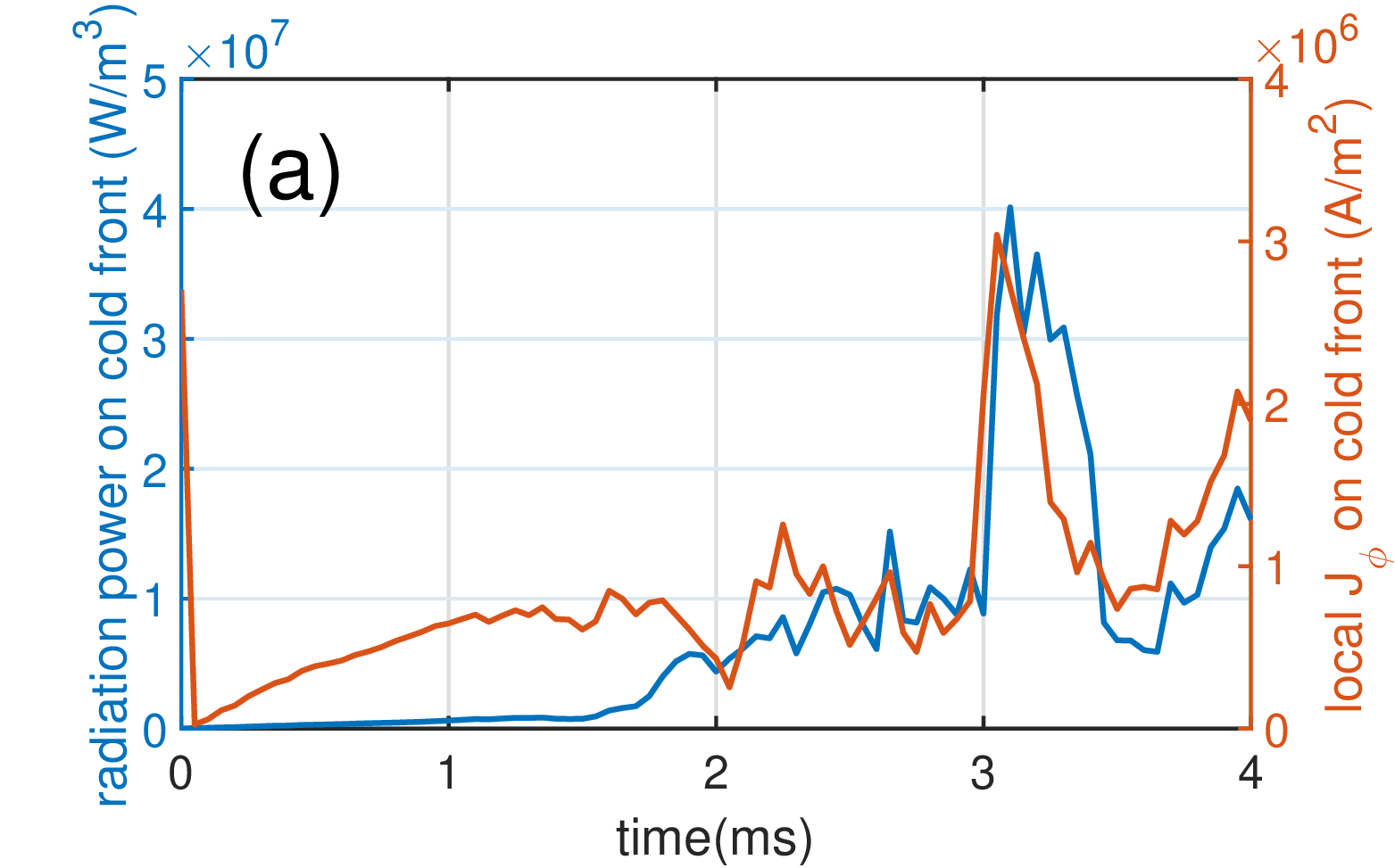}
			\includegraphics[width=0.85\textwidth,height=0.35\textheight]{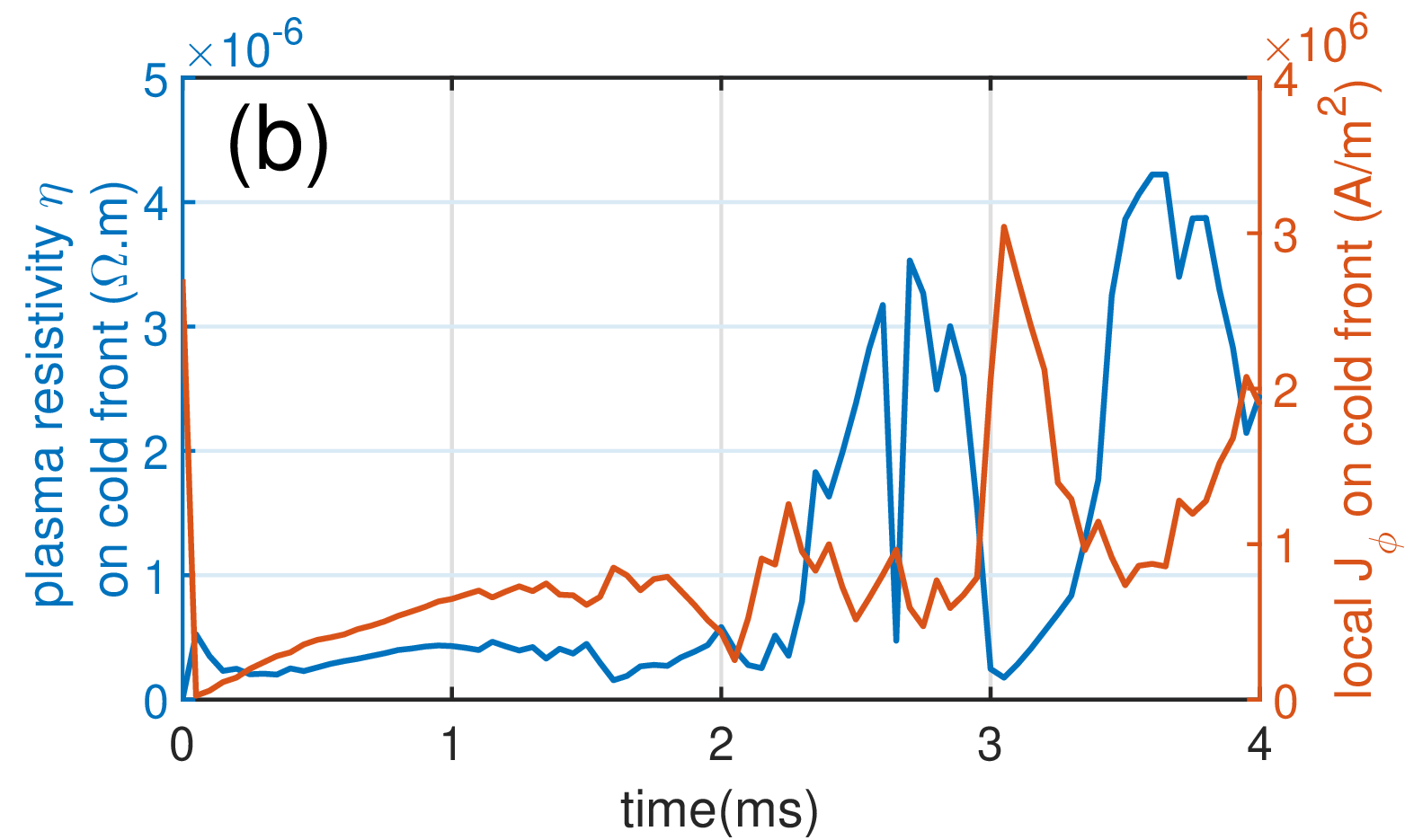}
		\end{center}
		\caption{(a) The local impurity radiation power (blue line) and toroidal current density $J_{\phi}$ (orange line), and (b) the plasma resistivity $\eta$ (blue line) and toroidal current density $J_{\phi}$ (orange line) on the impurity cold front as functions of time for the He impurity from the 3D simulation case.}
		\label{fig:3D-Jphi source}
	\end{figure}

	The source of the radial current perturbation $J_r$ is similar to the 2D case that the localized impurity radiation cooling modifies the local equilibrium and enforces a new local force balance, which can be seen from the variation of the surface-averaged current perturbation $J_r$ that is approximately balanced by the poloidal pressure gradient perturbation (Fig. \ref{fig:3D-Jr source}).
	A locally enhanced radiation power at the impurity cold front leads to a locally enhanced toroidal current perturbation $J_{\phi}$ (Fig. \ref{fig:3D-Jphi source}a). The impurity radiation modifies the current density mainly through the temperature dependent plasma resistivity, as indicated by the anti-correlation between the local resistivity and the current density variations in time (Fig. \ref{fig:3D-Jphi source}b).


	\section{Discussion and summary}
	\label{Sec:summary}
	
	\subsection{Discussion}
\begin{figure}[ht]
		\begin{center}
			\includegraphics[width=0.85\textwidth,height=0.35\textheight]{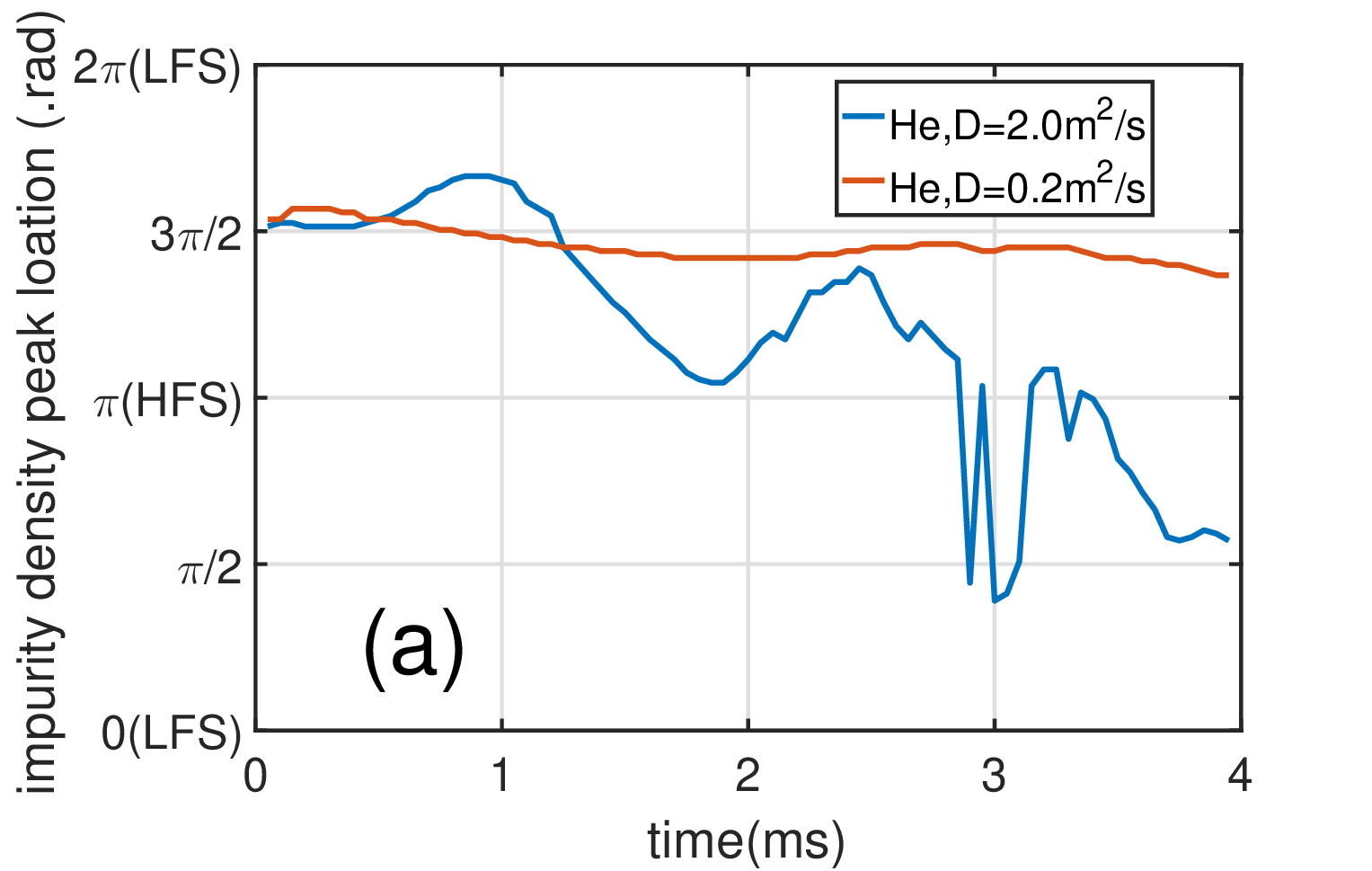}
			\includegraphics[width=0.85\textwidth,height=0.35\textheight]{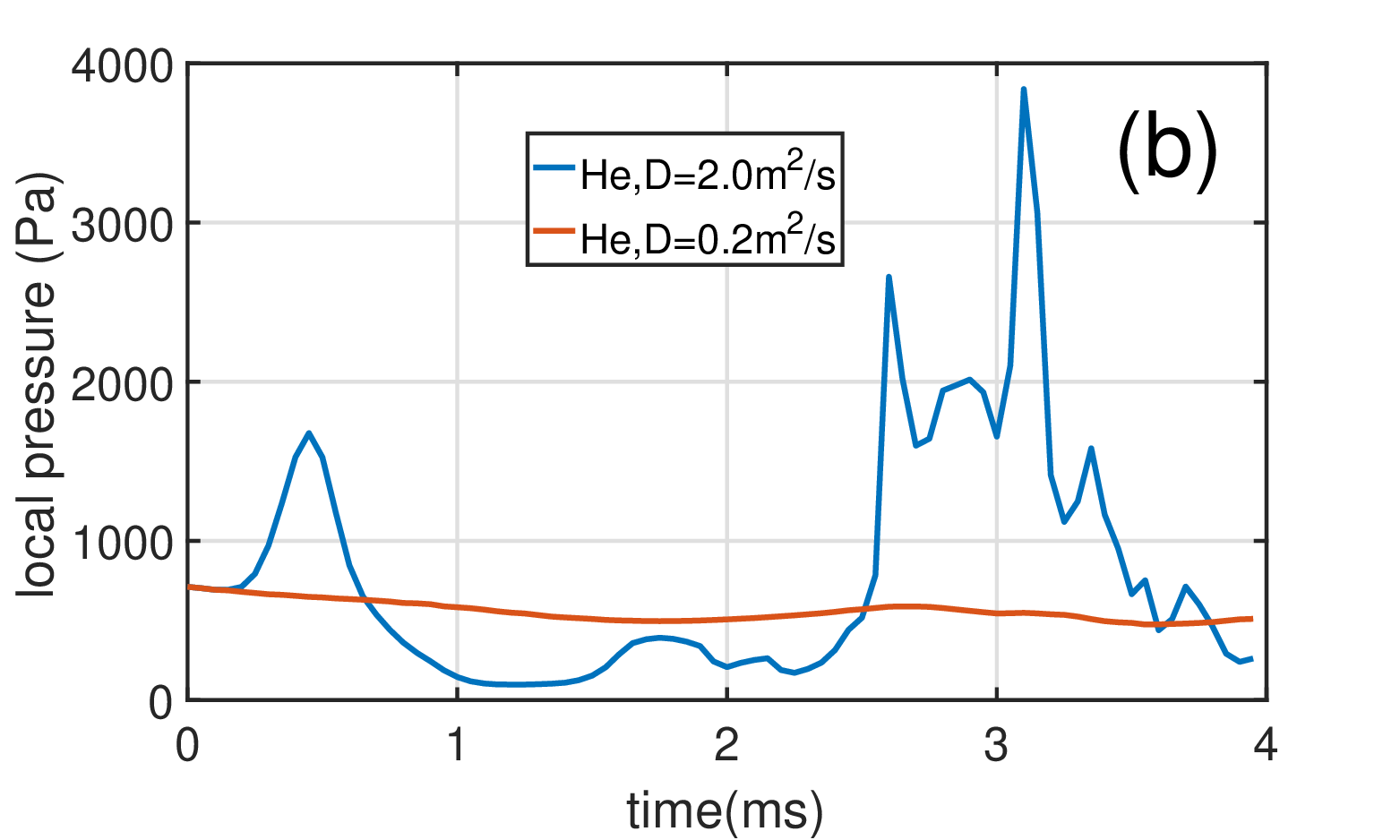}
		\end{center}
		\caption{(a) Impurity density peak poloidal angle in the injection poloidal plane and (b) local pressure $P$ at the bottom of the $q=2$ surface for the two He impurity cases with different diffusivity values as functions of time from 3D simulation cases.}
		\label{fig:3D-nd}
	\end{figure}

\begin{figure}[ht]
		\begin{center}
			\includegraphics[width=0.85\textwidth,height=0.5\textheight]{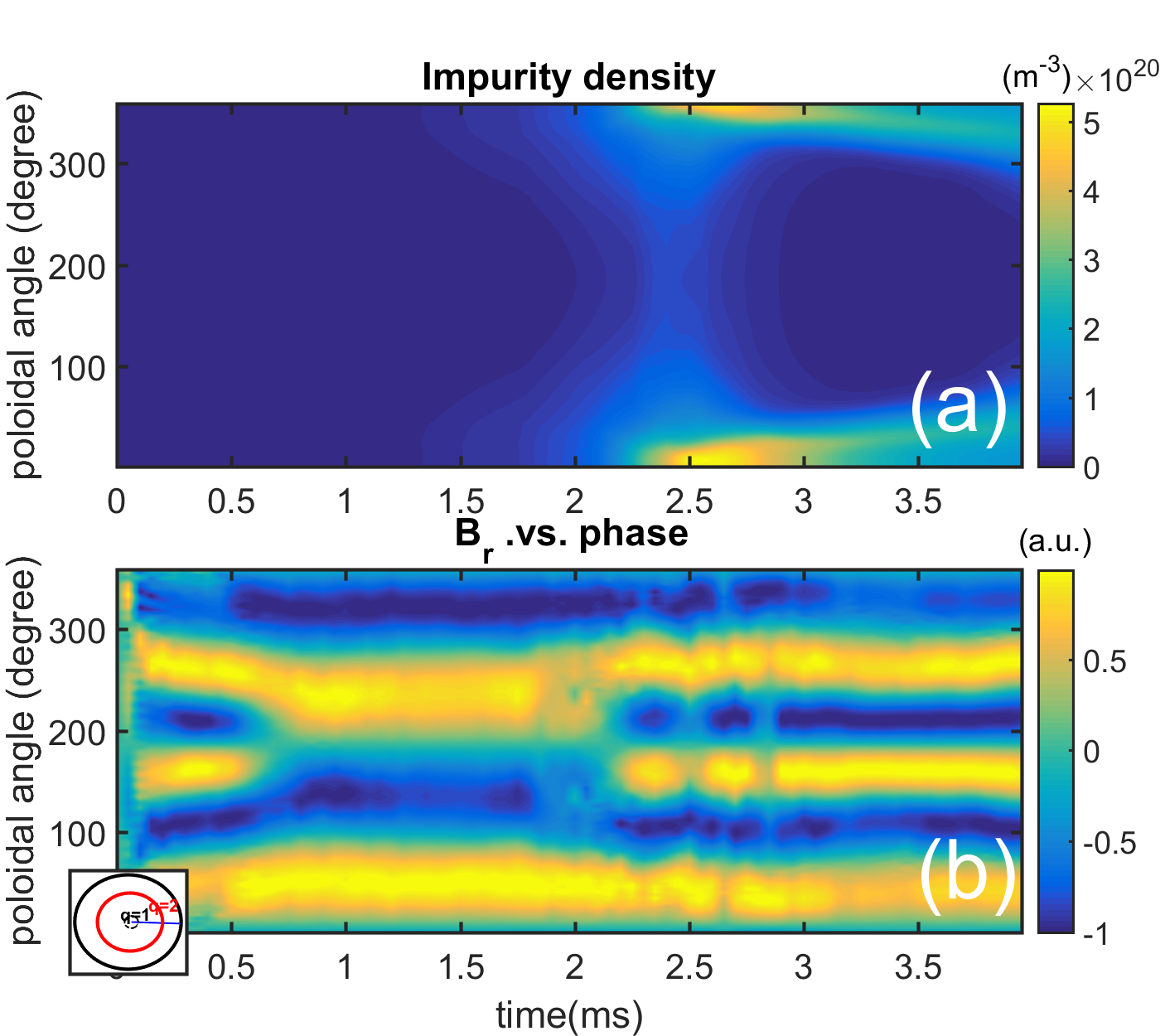}
		\end{center}
		\caption{Poloidal distributions of (a) the impurity density $n_{imp}(\theta) = \int_{0}^{a}n_{imp}(r,\theta,\phi=0) dr/a$, and (b) the normalized $n=1$ component of radial magnetic field $B_r$ as functions of time for He injection from the outer mid-plane.}
		\label{fig:3D-up dw sym}
	\end{figure}

	Although the poloidal reversal of the He impurity flow direction has been the focus of our study, there are other He impurity injection experiments with lower injector valve voltage where such a reversal is absent \cite{Li_2020}.
	Since the impurity source is static without injection velocity in the simulation and its initial penetration mainly comes from diffusion, a He injection case with much smaller diffusivity $D=0.2m^2/s$ is adopted to simulate the He injection experiment with lower injector valve voltage, and the simulation result is consistent with the experimental observations, where the impurity density only drift towards the HFS (Fig. \ref{fig:3D-nd}a). This result can also be understood from the local enhanced-$\beta$ plasmoid formation observed in the early stage of the faster He impurity injection case, but not found in the slower injection case of simulations (Fig. \ref{fig:3D-nd}b). Apparently, the impurity injection with a sufficient speed of penetration to the necessary temperature range is critical to a almost full ionization and the local enhanced-$\beta$ plasmoid formation, and hence the subsequent poloidal flow reversal. Another factor determining the drift direction of injected impurity density is known to be the poloidal location of the impurity injection.
	To further confirm this, we run a case with impurity injection from the outer mid-plane location ($\theta=0$), and indeed no impurity asymmetric poloidal drift flow or magnetic island rotation is found with this up-down symmetric injection setup (Fig. \ref{fig:3D-up dw sym}).


	In this work, the toroidal flow shows less correlation with the poloidal impurity density flow or the island rotation.
	Experiments on D\uppercase\expandafter{\romannumeral3}-D have demonstrated that the toroidal radiation asymmetries during the impurity injection are largely driven by the $n=1$ MHD modes \cite{Shiraki_2015}, and later experiments show that the $n=1$ MHD modes have little effects on the asymmetric evolution in the poloidal plane \cite{Eidietis2017pop}.
	Previous simulation works have demonstrated the helical spread of impurity plume that largely due to the helical magnetic field structure \cite{Izzo_2015,Zeng_2022}.
	In line with those findings, our work indicates that the dynamics of the impurity density and flow as well as the island rotation in the poloidal plane is dominated by the poloidal component of EM force and less relevant to the toroidal asymmetry. This may have certain implications on the optimal strategies for disruption mitigation and impurity assimilation efficiency, involving, e.g. the radiation poloidal peaking factor (PPF).


	\subsection{Summary}
	In this work, we have reproduced the dependence of the poloidal directions of impurity density drift and the magnetic island rotation on the impurity species observed in the J-TEXT experiment. A local enhanced-$\beta$ plasmoid formed at the early stage of the low-Z He impurity injection is found critical to the direction of the asymmetric impurity density poloidal flow. The electromagnetic force produced by the impurity radiation is shown as the main driver to the poloidal flow and the corresponding magnetic island rotation, which primarily comes from the toroidal symmetric components of perturbation.
	The penetration of the injected material is thus expected to be critical to both the fusion plasma fueling and the disruption mitigation, especially in the large scale fusion reactors, and further study is planned for future tokamak devices like ITER and CFETR.

	\section{Acknowledgments}
	We are grateful for the supports from the NIMROD team and the J-TEXT team. This work was supported by the National Magnetic Confinement Fusion Program of China (Grant No. 2019YFE03050004), the National Natural Science Foundation of China (Grant Nos. 51821005 and 12175228), Collaborative Innovation Program of Hefei Science Center, CAS (Grant No. 2021HSC-CIP007), and U.S. Department of Energy (Grant Nos. DE-FG02-86ER53218 and DE-SC0018001).

\newpage
\section{Reference}

\bibliographystyle{iopart-num}
\bibliography{eastexptm}


	\newpage

	\newpage

\newpage

	\newpage

	\newpage

	\newpage

	\newpage

	\newpage

	\newpage

	\newpage

	\newpage

	\newpage

	\newpage

	\newpage


\end{document}